\documentclass[sigconf,9pt]{acmart}

\usepackage[utf8]{inputenc}
\usepackage{tikz}
\usepackage{amsmath}
\usepackage{enumitem}
\usepackage{listings}
\usepackage{colortbl}
\usepackage{xcolor}
\usepackage{graphicx}
\usepackage{booktabs}
\usepackage{algorithm}
\usepackage{algpseudocode}
\usepackage{float}
\usepackage{adjustbox}
\usepackage{multirow}
\usepackage{pgfplots}
\usepackage{caption}
\usepackage{subcaption}
\usepackage[inkscapelatex=false]{svg}
\usepackage{svg}
\usepackage{amsmath}
\usepackage{wrapfig}
\usepackage{booktabs}
\usepackage{siunitx} 
\usepackage{tabularx}  
\usepackage{makecell}

\setlist[enumerate]{leftmargin=0.9cm}

\algrenewcommand\algorithmicrequire{\textbf{Input:}}
\algrenewcommand\algorithmicensure{\textbf{Output:}}
\newcommand{\fakepar}[1]{\vspace{1mm}\noindent\textbf{#1:}}
\newcommand{\fakeparNum}[1]{\vspace{1mm}\noindent\textbf{#1}}

\newcounter{boxlblcounter}  

\newcommand*\circled[1]{%
    \tikz[baseline=(char.base)]{%
        \node[shape=circle, fill, inner sep=2pt] (char) {\textcolor{white}{#1}};%
    }%
}

\definecolor{codegreen}{rgb}{0,0.6,0}
\definecolor{codegray}{rgb}{0.5,0.5,0.5}
\definecolor{codepurple}{rgb}{0.58,0,0.82}
\definecolor{backcolour}{rgb}{0.95,0.95,0.92}

\definecolor{shellbg}{gray}{0.95} 
\definecolor{shelltext}{rgb}{0,0,0}
\definecolor{shellstring}{rgb}{0.0,0.0,0.0}

\lstdefinelanguage{shell}{
    sensitive=false, 
    morecomment=[l]{\$}, 
    morecomment=[s]{/*}{*/}, 
    morestring=[b]" 
} %
\lstdefinestyle{shellst}{  
    backgroundcolor=\color{shellbg},   
    commentstyle=\color{shelltext}\bfseries,
    numberstyle=\tiny\color{codegray},
    stringstyle=\color{shellstring},
    basicstyle=\ttfamily,      
    frame=single,
    breaklines=true,                 
    captionpos=b,                    
    keepspaces=true,                 
    numbers=left,                    
    numbersep=10pt,                  
    showspaces=false,                
    showstringspaces=false,
    showtabs=false,
    columns=flexible,
}

\lstdefinestyle{codest}{
    backgroundcolor=\color{shellbg},   
    commentstyle=\itshape\color{purple!40!black},
    keywordstyle=\bfseries\color{green!40!black},
    numberstyle=\tiny\color{codegray},
    stringstyle=\color{magenta},
    basicstyle=\ttfamily\small,
    identifierstyle=\color{blue},
    breakatwhitespace=false,
    breaklines=true,                 
    captionpos=b,                    
    keepspaces=true,                 
    numbers=left,                    
    numbersep=5pt,                  
    showspaces=false,                
    showstringspaces=false,
    showtabs=false,                  
    tabsize=2
}

\usepackage{xspace}

\newcommand{\tool}{CheckMate\xspace}

\AtBeginDocument{%
  \providecommand\BibTeX{{%
    \normalfont B\kern-0.5em{\scshape i\kern-0.25em b}\kern-0.8em\TeX}}}

\newcommand{\needo}[1]{\textcolor{black}{#1}}
\newcommand{\shepherdchange}[1]{\textcolor{black}{#1}}

\begin{document}

\title{CheckMate:\\ LLM-Powered Approximate Intermittent Computing}

\author{Abdur-Rahman Ibrahim Sayyid-Ali}
\email{25100204@lums.edu.pk}
\affiliation{
  \institution{LUMS}
  \country{Pakistan}
}

\author{Abdul Rafay}
\email{abdul.rafay@lums.edu.pk}
\affiliation{
  \institution{LUMS}
  \country{Pakistan}
}

\author{Muhammad Abdullah Soomro}
\email{msoomro@umass.edu}
\affiliation{
  \institution{UMass Amherst}
  \country{United States}
}

\author{Muhammad Hamad Alizai}
\email{hamad.alizai@lums.edu.pk}
\affiliation{
  \institution{LUMS}
  \country{Pakistan}
}

\author{Naveed Anwar Bhatti}
\email{naveed.bhatti@lums.edu.pk}
\affiliation{
  \institution{LUMS}
  \country{Pakistan}
}

\renewcommand{\shortauthors}{Sayyid-Ali, et al.}

\begin{abstract}
Batteryless IoT systems face energy constraints exacerbated by checkpointing overhead. Approximate computing offers solutions but demands manual expertise, limiting scalability. This paper presents CheckMate, an automated framework leveraging LLMs for context-aware code approximations. \tool\ integrates validation of LLM-generated approximations to ensure correct execution and employs Bayesian optimization to fine-tune approximation parameters autonomously, eliminating the need for developer input. Tested across six IoT applications, it reduces power cycles by up to 60\% with an accuracy loss of just 8\%, outperforming semi-automated tools like ACCEPT in speedup and accuracy. CheckMate’s results establish it as a robust, user-friendly tool and a foundational step toward automated approximation frameworks for intermittent computing.
\end{abstract}

\maketitle

\section{Introduction}
\label{sec:intro}

Energy harvesting enables batteryless IoT devices to operate without regular batteries, reducing maintenance costs and supporting multi-year unattended deployments~\cite{afanasov2020battery}. However, environmental energy sources are often erratic, leading to frequent and unpredictable power failures, emphasizing the critical importance of energy efficiency in batteryless IoT applications~\cite{lucia2017intermittent,jia2022transient,ahmed2021survey, ahmed2024internet}.

Traditionally, checkpointing methods have served as the cornerstone for managing program state across power failures in energy-harvesting devices, despite introducing considerable overhead~\cite{mementos,bhatti2016efficient,bhatti2017harvos,jayakumar2015quickrecall,alharbi2023checkpointing}. This computing paradigm, which frequently saves and restores program state, is commonly referred to as \emph{intermittent computing}.
Over the past decade, researchers have sought to mitigate this overhead through various innovations, such as optimizing checkpoint placement~\cite{bhatti2017harvos,ahmed2019efficient,maeng2018adaptive}, virtualizing memory~\cite{maioli2021alfred}, fragmenting tasks into idempotent code blocks~\cite{colin2016chain,maeng2017alpaca}, and employing event-driven programming models~\cite{yildirim2018ink}. Despite these efforts, the marginal gains from further optimization are diminishing, signaling that the traditional focus on reducing checkpointing overhead in intermittent computing has reached its practical limits~\cite {NotToCheckpoint}.


In this work, we shift the paradigm from minimizing checkpoint overhead to directly reducing the energy consumed by computations.
Leveraging approximate computing techniques, we relax computational accuracy in targeted ways, which lowers energy demands and enables more tasks to be completed per power cycle\footnote{A power cycle refers to a single cycle of intermittent computing during which the system is powered \textit{ON} and performs computations}, and reduces the need for frequent checkpoints. This strategy is particularly advantageous for intermittent computing on batteryless IoT systems, where energy availability is unpredictable and resources are limited. In such environments, slight compromises in accuracy are acceptable to achieve task completion with reasonable precision~\cite{lin2023intermittent,bambusi2022case}, making approximate computing a suitable solution for applications such as filtering, machine learning, image processing, and data fusion. These areas are essential to many IoT systems, where efficient computation is critical to maintaining functionality under extreme energy constraints.

\fakepar{Challenges}
While several researchers have already introduced approximation techniques in batteryless systems, signaling the early sparks of this paradigm shift, the path to broader adoption remains obstructed. The absence of an automated framework for efficiently implementing these approximations stands as the primary challenge. Researchers have been limited to optimizing approximation techniques for specific applications and scenarios, resulting in studies that are replicable only within those exact contexts~\cite{javed2023moptic,bambusi2022case,ganesan2019s,islam2019zygarde,10.1145/3581791.3596845,9923863,10.1145/3607918,lin2023intermittent,lucaSenSys25}. At its core, the challenge revolves around transforming the abstract potential of approximation into practical solutions that work across diverse applications. Developers face a fundamental dilemma: how to achieve the right balance between energy savings and computational precision without undermining the reliability of program output. We have distilled this challenge into three key areas that require attention to advance the field.

\fakeparNum{}\circled{1} Identifying suitable opportunities for approximation requires more than just technical know-how; it demands a nuanced understanding of the application's semantics and execution context, hardware architecture, and the particularities of various approximation strategies. Poor judgment in this process can compromise energy efficiency or, worse, result in inaccuracies that degrade system performance. Developers must navigate this terrain carefully, ensuring that the energy-accuracy trade-offs they introduce are both purposeful and effective.

\fakeparNum{}\circled{2} Even when approximation opportunities are identified, the next hurdle lies in the painstaking process of manual tuning. This step is often labor-intensive, requiring developers to carefully tailor adjustments to match the specific needs of each application. This fine-tuning process, while essential, is prone to errors and demands significant time and effort. Developers must walk a tightrope, ensuring that the savings in energy consumption do not come at the expense of unacceptable losses in computational fidelity.

\fakeparNum{}\circled{3} The challenges do not end with implementation. Even the most thoughtfully designed approximations must withstand the unpredictable realities of real-world deployment. Predicting how these approximations will behave post-deployment, under varying conditions and across multiple use cases, introduces yet another layer of complexity. Without a way to accurately foresee the impact of approximations, developers risk deploying systems that fail to meet performance expectations, compromising both scalability and reliability.


\fakepar{Contribution} 
We introduce \tool\footnote{The name `CheckMate' embodies the tool's core functionality, automatically integrating \textit{CHECK}pointing and approxi\textit{MATE} computing techniques. Source code is publicly 
available at \shepherdchange{https://github.com/SYSNET-LUMS/CheckMate}}, an open-source novel automated framework that harnesses the intelligence of Large Language Models (LLMs) to effectively balance energy consumption and computational accuracy in batteryless IoT systems. Unlike existing frameworks constrained by laborious manual tuning \cite{accept} or simplistic one-size-fits-all strategies~\cite{rumba}, \tool\ breathes sophistication into the process by merging LLM-driven insights with automated validation and optimization mechanisms. LLMs' semantic understanding and contextual awareness enable our approach to identify effective points for approximation, determine suitable techniques, and specify their application.
However, today's LLMs have inherent limitations: they cannot guarantee that the modified code for embedded platforms compiles correctly, assess the energy requirements of the target platform to estimate power cycles, or ensure an efficient trade-off between energy efficiency and computational accuracy. 
To address these challenges, \tool\ integrates a rigorous validation process to verify the correctness of LLM-generated code. This process detects compile-time and runtime errors, iteratively feeding them back to the LLM for correction until the code functions as intended. Once validated, the code is subjected to fine-tuning through Bayesian optimization, a probabilistic model-based approach, within a cycle-accurate simulator designed for intermittent computing. This optimization identifies the balance point that minimizes power cycles while ensuring computational accuracy remains within user-defined error bounds. This whole process is depicted in Figure~\ref{fig:flow} and described in Section~\ref{sec:design-choices}.

\fakepar{Benefits}
The effectiveness of CheckMate is validated through a comprehensive evaluation encompassing six diverse IoT applications, five energy harvesting traces, and both simulation and real-world scenarios. CheckMate achieved significant energy efficiency, reducing power cycles by 15–60\% while maintaining error rates within the range of 6–25\%. Compared to semi-automated frameworks like ACCEPT~\cite{accept}, CheckMate demonstrates comparable or superior performance while eliminating the need for expert intervention. In real-world hardware evaluations, results closely align with simulation predictions, reinforcing the tool’s reliability. Additionally, a user study involving computer science students highlights its practicality, with a 5x reduction in time to apply approximations and a significant decrease in error rates compared to manual methods. This remarkable performance was attained through a seamless \textit{one-click} process, which obviates the need for the labor-intensive fine-tuning typically required by existing frameworks. 

As developers increasingly seek ways to optimize batteryless applications, \tool\ offers a glimpse into the future; where intelligent automation, backed by LLMs, transforms the approach to approximate intermitted computing for batteryless IoT.

\label{sec:checkmate}

\section{Design Choices and Workflow}

\label{sec:design-choices}
The development of \tool\ introduced several challenges that prompted iterative design decisions. In the following, we outline the key design choices and lessons learned through micro-experiments and continuous refinements.

\subsection{Error-resolution feedback loop}

In the initial stages of CheckMate's development, we began with a simple experiment: entrusting the LLM, with its purported magic-like powers, to approximate the code for execution on a batteryless platform. The objective was for the LLM to process the entire codebase in one submission and return the approximated version. While the LLM demonstrated a rudimentary ability for code approximation, it frequently produced code riddled with compilation or runtime errors.
The generated code required manual adjustments for seamless integration, which contradicted our objective to minimize developer intervention. This outcome highlighted the need for a robust error-resolution mechanism within \tool.

\fakepar{Approach}
We implemented an iterative feedback mechanism for \tool to effectively mitigate errors introduced by LLM-generated code. Each iteration involved interaction between the LLM and \tool's external script, which compiled and executed the LLM-generated approximated code. Compilation failures and runtime errors were systematically logged and fed back to the LLM, creating a cycle of autonomous refinements and correction. This iterative process marks a foundational step towards achieving a one-click solution, as it eliminates the need for manual debugging.

\subsection{Dynamic tuning via adjustable knobs}
A key challenge was enabling the LLM to understand the impact of its approximations on both output accuracy and energy efficiency. Today's LLMs cannot measure hardware energy consumption or assess the resulting trade-offs between output accuracy and energy usage. This limitation necessitated the integration of an external simulator to provide the essential feedback needed for informed decision-making.

Building on the initial error-resolution feedback loop, we introduced a secondary feedback mechanism leveraging a cycle-accurate simulator. 
The application is executed inside this simulated environment using user-provided input traces. The simulator evaluates output error rates and power cycle reductions, producing quantitative metrics that the LLM itself cannot compute. This approach bridges the gap between the LLM’s capabilities and the requirements for optimizing approximations.

However, the fine-tuning process did not proceed as anticipated, frequently oscillating between overly conservative and overly aggressive adjustments. This behavior complicates the achievement of an effective balance and prolongs feedback cycles, increasing both the cost of LLM interactions and the computational workload of running the simulator. We needed a fine-grained exploration of the LLM-suggested approximation techniques and their impact on errors and energy consumption, a task that LLMs were ill-suited for due to their inefficiency in handling such domain-specific optimization challenges.

\fakepar{Approach}
We significantly revised our approach to the fine-tuning process by redefining the role of the LLM. Instead of relying on it to produce hardcoded approximation parameters, we prompted it to introduce adjustable \emph{knobs} within the code to control approximation levels, offloading unnecessary burdens from the LLM.

For each approximation opportunity, such as the truncation factor in loop perforation, the LLM was guided to embed these knobs (variables) into the codebase along with their suggested ranges. This design enables precise control over the degree of approximation for specific functions, allowing targeted tuning to adhere to defined error tolerances. 

To automate the process of identifying the optimal knob values, we employed Bayesian optimization, which determines knob configurations that achieve the desired balance between energy efficiency and computational accuracy, accounting for the interactions among multiple knobs. 
\shepherdchange{The knobs also serve as mechanisms to mitigate inaccurate approximations by enabling the optimization algorithm to `tune out'\footnote{\shepherdchange{The optimization algorithm can adjust the knob to a value that ensures the corresponding code section behaves as it would in the original, unmodified version, effectively nullifying the approximation.}} the approximation.}  

\subsection{Context-aware Chain-of-Thought}

Another challenge was the tendency of LLMs to miss critical approximation opportunities or apply suboptimal approximations, often leading to an unnecessary increase in the number of knobs leading to a substantial increase in the runtime of Bayesian optimization without proportional benefits. This issue primarily stemmed from the LLM's limited ability to holistically analyze the code structure when processing an entire codebase in one step, restricted by inference time and token generation limits. Furthermore, analyzing the entire codebase disrupted the LLM’s tracking of the execution sequence of different functions in the code, which is essential for applying effective approximations. This context loss frequently led to \emph{hallucinations}, where the LLM generated irrelevant or incorrect outputs.

\fakepar{Approach} 
To address these issues, multiple strategies are available, including \textit{chain-of-thought} (CoT) prompting with few-shot learning, fine-tuning models with additional data, or implementing retrieval-augmented generation (RAG) to contextualize LLM outputs. 
\shepherdchange{Due to the absence of sufficiently large and relevant datasets on approximate computing, running a specialized fine-tuned model or one equipped with retrieval-based generation is impractical.}  
Instead, we adopt a context-aware CoT strategy with few-shot learning to enhance the quality of the LLM’s code output. \shepherdchange{This approach allows us to supplement the LLM’s inherent reasoning with explicit problem-solving steps that align with expert decision-making, guiding the model toward more accurate and contextually appropriate outputs. Additionally, incorporating a few carefully selected examples of approximations helps the LLM recognize how knobs and approximation techniques should be structured within the code}

The strategy begins with \textit{purpose identification}, where the LLM determines the application's primary objective and provides summaries for each function. This step establishes a high-level understanding of the codebase context, which is logged for later use. The LLM then generates a list of functions suitable for approximation, serving as a filter to avoid suboptimal modifications, ensuring that the LLM performs context-aware function selection, providing a foundation for targeted approximations.
As the process continues, the conversation history may exceed the LLM’s context window, leading to a loss of initial context. To mitigate this, subsequent interactions occur in a new conversation, where the LLM is reintroduced to the codebase using only the previously generated summaries and the list of functions selected for approximation.

We iterate over the selected functions using a function call graph (FCG) to guide the process in a sequential order. For each selected function, the LLM performs two tasks. First, it identifies potential approximation strategies and reasons for their impact on the codebase. Next, based on this reasoning, the LLM is tasked with implementing the most suitable approximation for the selected functions. This context-aware strategy enables precise and effective approximations tailored to the application’s operational requirements.

\begin{figure}[tb]
    \centering
    \includegraphics[width=\columnwidth]{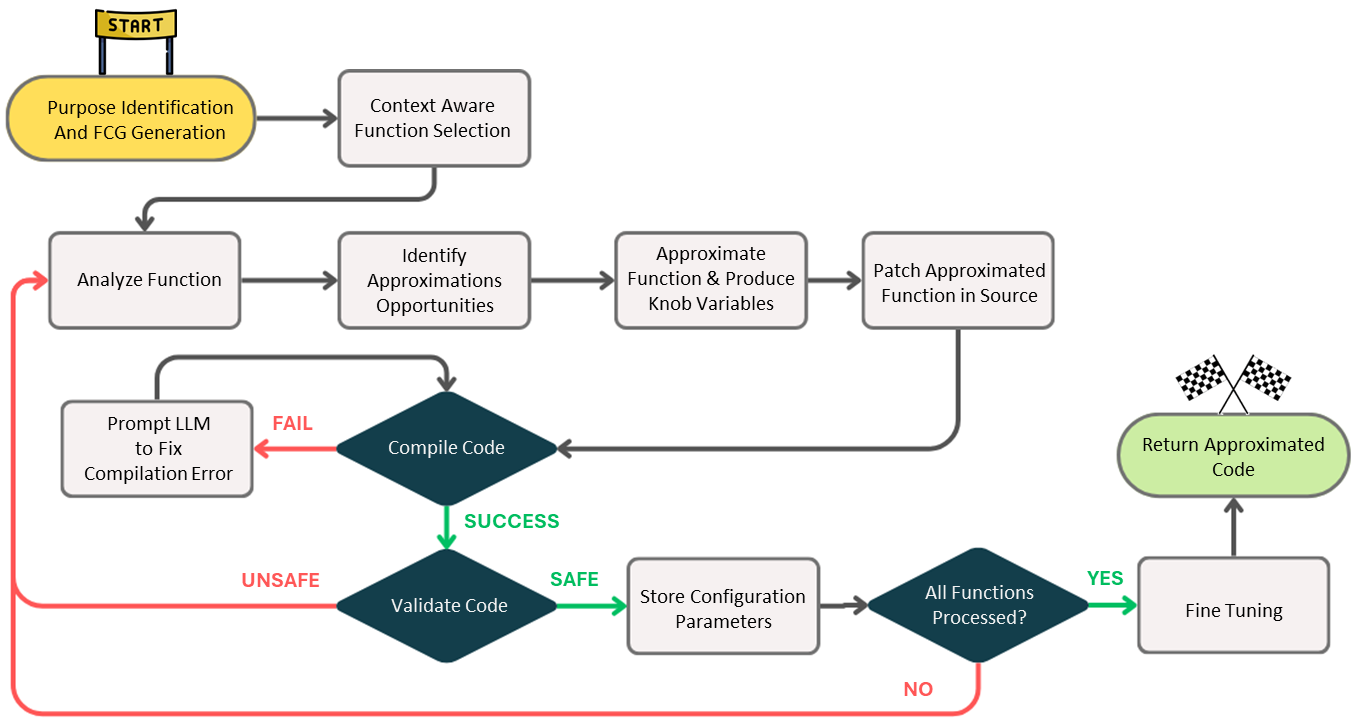}
    \caption{\tool workflow.}
    \label{fig:flow}
\end{figure}

\subsection{Workflow}

Each design choice in developing \tool\ stemmed from addressing practical challenges and was iteratively refined through experimentation. The integration of \textit{feedback loops}, the deployment of \textit{knobs}, the use of \emph{Bayesian optimization}, and the adoption of a \textit{CoT} prompting strategy collectively enhance \tool's ability to generate effective approximations \shepherdchange{while minimizing errors}, for batteryless IoT applications through a one-click solution.

Figure~\ref{fig:flow} presents a detailed view of \tool’s workflow, illustrating the seamless interaction between its components. It outlines each step, from function identification to fine-tuning, showcasing the cohesive and systematic flow that underpins the approximation process.


\section{Architecture and Implementation}
\definecolor{mGreen}{rgb}{0,0.6,0}
\definecolor{mGray}{rgb}{0.5,0.5,0.5}
\definecolor{mPurple}{rgb}{0.58,0,0.82}
\definecolor{backgroundColour}{rgb}{0.95,0.95,0.92}

\lstdefinestyle{CStyle}{
    backgroundcolor=\color{backgroundColour},   
    commentstyle=\color{mGreen},
    keywordstyle=\color{magenta},
    numberstyle=\tiny\color{mGray},
    stringstyle=\color{mPurple},
    basicstyle=\footnotesize,
    breakatwhitespace=false,         
    breaklines=true,                 
    captionpos=t,                    
    keepspaces=true,                 
    numbers=left,                    
    numbersep=5pt,                  
    showspaces=false,                
    showstringspaces=false,
    showtabs=false,                  
    tabsize=2,
    language=C
}
\lstset{
    belowskip=10pt
}
\lstdefinelanguage{CustomLang}{
    basicstyle=\ttfamily\small, 
    backgroundcolor=\color{gray!10}, 
    frame=single, 
    rulecolor=\color{black}, 
    breaklines=true, 
    captionpos=b, 
    numbers=none, 
    keywordstyle=\bfseries, 
}

\lstdefinelanguage{customc}{
    sensitive=true,
    keywords={auto,break,case,char,const,continue,default,do,double,%
        else,enum,extern,float,for,goto,if,int,long,register,%
        return,short,signed,sizeof,static,struct,switch,typedef,%
        union,unsigned,void,volatile,while},
    morekeywords=[2]{printf,scanf,main},
    morecomment=[l]{//},
    morecomment=[s]{/*}{*/},
    morestring=[b]",
    morestring=[b]',
    keywordstyle=\color{blue!70!black}\bfseries,
    keywordstyle=[2]\color{purple!70!black}\bfseries,
    commentstyle=\color{green!50!black}\itshape,
    stringstyle=\color{red!70!black},
    basicstyle=\normalsize\ttfamily,
    numbers=left,
    numberstyle=\normalsize,
    stepnumber=1,
    numbersep=8pt,
    frame=lines,
    backgroundcolor=\color{gray!5},
    xleftmargin=15pt,
    xrightmargin=5pt,
    captionpos=b,
    linewidth=\columnwidth,  
    breaklines=true
}

\lstdefinelanguage{customjson}{
    sensitive=true,
    keywords={true,false,null},
    showstringspaces=false,
    morecomment=[l]{//},
    morecomment=[s]{/*}{*/},
    morestring=[b]",
    keywordstyle=\color{blue!70!black}\bfseries,
    commentstyle=\color{green!50!black}\itshape,
    stringstyle=\ttfamily\color{red!70!black},
    numberstyle=\color{magenta},
    basicstyle=\normalsize\ttfamily,
    numbers=left,
    numberstyle=\normalsize,
    stepnumber=1,
    numbersep=8pt,
    frame=lines,
    backgroundcolor=\color{gray!5},
    xleftmargin=15pt,
    xrightmargin=5pt,
    captionpos=b,
    linewidth=\columnwidth,
    breaklines=false,
    breakindent=0pt,
    postbreak=\mbox{\textcolor{red}{\ensuremath{\hookrightarrow}}},
    identifierstyle=\color{blue!70!black},
    literate=
     *{0}{{{\color{magenta}0}}}{1}
      {1}{{{\color{magenta}1}}}{1}
      {2}{{{\color{magenta}2}}}{1}
      {3}{{{\color{magenta}3}}}{1}
      {4}{{{\color{magenta}4}}}{1}
      {5}{{{\color{magenta}5}}}{1}
      {6}{{{\color{magenta}6}}}{1}
      {7}{{{\color{magenta}7}}}{1}
      {8}{{{\color{magenta}8}}}{1}
      {9}{{{\color{magenta}9}}}{1}
}

\begin{figure*}
    \centering
    \includegraphics[width=\textwidth]{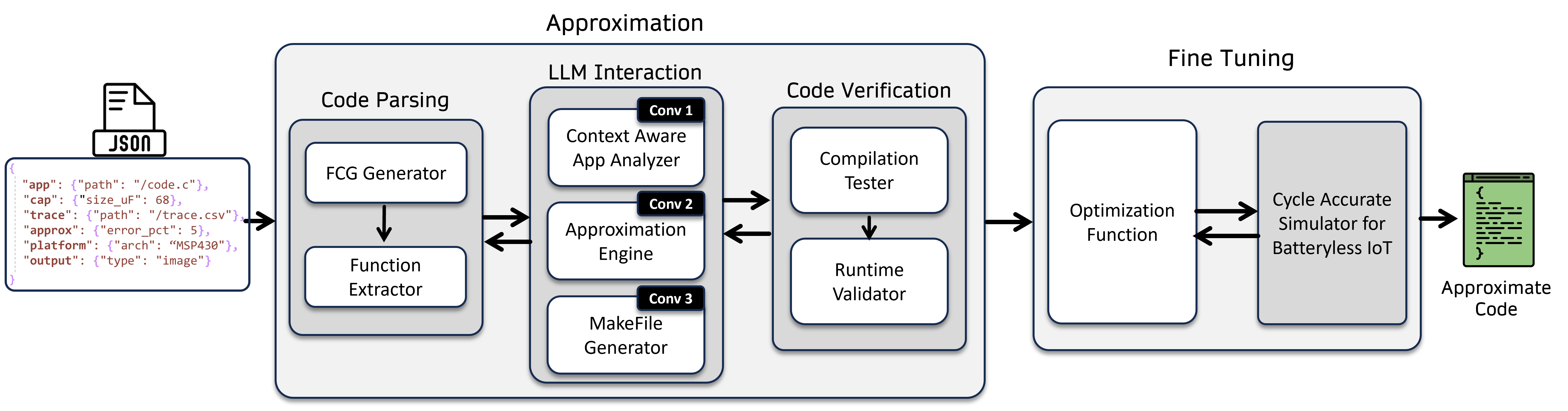}
        \caption{\tool\ architecture.}

    \label{fig:archi}
\end{figure*}


\tool\ architecture enables seamless interaction among its core components, ultimately transforming user input into approximate code, as illustrated in Figure \ref{fig:archi}. In the following, we provide an in-depth exploration of each module’s implementation.

\subsection{User Input}

\tool\ requires six essential inputs in JSON format, which users may modify: original application source-code, used to identify approximation opportunities; \textit{input traces}, representing potential application inputs and energy harvesting traces, or, if unavailable, detailed input characteristics to simulate a trace; \textit{accuracy class ($a_c$)}, used to calculate output deviations (cf. Section \ref{sec:fine-tuning} and Table\ref{tab:accuracy-classes}); and \textit{error bound ($e_b$)}, which defines acceptable deviation limits in the output. Additionally, knowledge of the \textit{platform} and its architecture (e.g., Arm Cortex M or MSP430) aids in precise energy consumption estimation, while the \textit{capacitor size} determines the minimum energy required to prevent non-progressive states, where the program fails to advance beyond a certain code segment~\cite{hester2017flicker}. \shepherdchange{An optional parameter, \textit{LLM temperature}, can be adjusted, with a default low value (0–0.3), as this range minimizes the LLM's randomness and is generally preferred for coding tasks~\cite{chen2021codex}.}

\subsection{Approximation}

After gathering the required inputs, \tool\ initiates the code approximation process. It begins by parsing the source code (Listing \ref{original-code}) and iteratively approximating functions selected by the LLM. Each approximated function undergoes verification to ensure it is error-free and safe for execution. We illustrate this workflow using the Sobel filter~\cite{kanopoulos1988design}, a common edge detection algorithm in IoT, as a running example.

\begin{lstlisting}[language=customc,caption={Original sobel filter code.}, label={original-code},  basicstyle=\small\ttfamily, numbers=none]
void sobel_filter(double* image) {

    image = load_image(IMAGE_SIZE);
    double* new_image = [IMAGE_SIZE];
    
    for (int i = 0; i < IMAGE_SIZE; i++) {
        /* Applying Sobel Filter */
    }
    send_to_classifier(new_image);
}

\end{lstlisting}

\subsubsection{Code parsing}
\label{sec:parsing}

It consists of two primary steps: generating the FCG and extracting function definitions.

\fakepar{FCG generation}
We employed Egypt~\cite{egypt} to generate the FCG, utilizing \texttt{gcc} to produce \texttt{.expanded} files for source code analysis and \texttt{Graphviz} for visualization. The resulting visualization file is processed to create an adjacency matrix representing the FCG.

To handle cyclic dependencies, cycles are arbitrarily removed~\cite{mohammed2024enabling}. After breaking these cycles, a topological sorting algorithm establishes the order for function approximation.

\fakepar{Function extraction}
We use Language Server Protocol (LSP)~\cite{microsoft2024lsp} to accurately locate the line and character positions of function definitions within the original codebase. A custom Python script extracts these functions and stores their content in a JSON file.

Together, these processes provide the functions, their dependencies, and the order in which functions are approximated (from least to most dependent), forming a robust foundation for the next stages in the pipeline.

\subsubsection{LLM interaction}

We utilized LangChain~\cite{langchain2024}, an open-source development framework, to streamline interactions with LLMs. LangChain's modular architecture supports integration with various LLMs, whether locally hosted or accessed through commercial APIs. For our evaluation, we employed OpenAI's GPT-4o, GPT-4o-mini, and Claude Sonnet 3.5 APIs~\cite{openai2024, anthropic_claude35}, while maintaining flexibility for future integration with other LLMs\footnote{We also intended to evaluate GPT-4o1; however, it remains in preview mode, and its API is not yet publicly available.}.

\begin{lstlisting}[language=customjson, caption={Function summaries and classification.}, label={lst:function-selection}, basicstyle=\small\ttfamily, numbers=none, breaklines=true, breakatwhitespace=false, columns=flexible, frame=single, linewidth=\columnwidth],
{
  code_summary: {
    "sobel_filter": "Applies sobel filter to image by...",
    "load_image": "Loads and preprocess the image to...",
    "send_to_classifier": "Uses BLE module to send...",
  },
  target_functions: {
    "sobel_filter": "approximate",
    "load_image": "do not approximate",
    "send_to_classifier": "do not approximate",
  }
}
\end{lstlisting}

All interactions with the LLM start with a \emph{system prompt} establishes the context, ensuring that the LLM's outputs align with the specific objectives and constraints of the task.
\needo{For certain steps} LLM begins with few-shot learning containing examples of original and approximated code for intermittent computing.

\fakepar{Function selection} Identifying functions for approximation is a critical step in our optimization process (\textbf{Conv 1}, Figure~\ref{fig:archi}). This is achieved through a structured sequence of prompts.

Initially, \needo {the LLM is passed a system prompt detailing its purpose and overall objectives [A.1.\ref{append:conv1sys}]. Next we present the entire application code to LLM to help it} understand the overall objective of the application. It then generates concise summaries for each function, describing their purpose, inputs, and outputs \needo{[A.1.\ref{append:conv1prompt1}]}.
Subsequently, each function is analyzed individually \needo{[A.1.\ref{append:conv1prompt2}]}. The LLM evaluates the role of each function within the application, determining which are suitable for approximation and identifying the potential benefits of such approximations.
Finally, using a dedicated prompt, the LLM classifies each function as either ``approximate'' or ``do not approximate'' (Listing \ref{lst:function-selection})\needo{[A.1.\ref{append:conv1prompt3}]}. This systematic approach allows the \tool to ensure that needless approximations and the resulting knobs are filtered out during the initial steps, as described next.

\begin{lstlisting}[language=CustomLang,caption={LLM suggested approximation.}, label={planning-step-example},  basicstyle=\small\ttfamily] 
Reducing Image Size

There are two ways to approximate by resizing:

1. To approximate code, reducing the `IMAGE_SIZE` variable is perforated, as it decreases the number of iterations in the Sobel filter loop, `load_image`, and `send_to_classifier` functions. However, this may introduce an input error in `send_to_classifier` if the image size no longer aligns with the model's requirements. In contrast, perforating the Sobel filter loop only reduces iterations within that loop by skipping or truncating some pixels, keeping the overall image size constant, but it provides less cycle reduction compared to reducing the image size.

2. ...
\end{lstlisting}

\fakepar{Approximating functions}  
Using the topological sort from the earlier steps and the list of functions selected for approximation, we filter the sort to include only the selected functions, which are processed sequentially. \needo{As shown in Figure~\ref{fig:archi}, this step occurs in a separate conversation with a distinct purpose. Lacking any prior context, the LLM is reintroduced to the codebase using only the brief summary from the previous discussion. Additionally, a new system prompt is used to align with the conversation's specific objective [A.2.\ref{append:conv2sys}].} 

The approximation process comprises two phases. In the first phase, the LLM is prompted to enumerate all possible approximations for the selected function and their potential benefits \needo{[A.2.\ref{append:conv2prompt1}]}. The LLM identifies suitable areas for approximation, highlights knobs, and returns a list of feasible code modifications (Listing~\ref{planning-step-example}) along with implementation details. As outlined in \textbf{Conv 2} (Figure~\ref{fig:archi}), this phase guides the LLM to identify and plan the most appropriate approximations for the application.

In the second phase, the LLM is prompted to implement the selected approximations, including creating knobs \needo{[A.2.\ref{append:conv2prompt2}]}. The LLM also defines the range of these knobs to facilitate later fine-tuning (cf. Section~\ref{sec:fine-tuning}). \needo{Depending on the potential approximation discussed in the previous response from the LLM, a few-shot example set may be provided to illustrate safe practices for applying specific types of approximations [A.2.\ref{append:fewshot-loop-perf}, A.2.\ref{append:fewshot-precision}].} As shown in the example in Listing~\ref{approximated-snippet}, the LLM applies loop perforation and introduces a knob. Up to this point, the interaction with LLM relies on plain text with embedded code snippets, but the variability in responses can complicate extracting the necessary information. To address this, a final prompt consolidates all data into a structured JSON object. The resulting JSON object (Listing~\ref{lst:json_example}) provides a concise summary of the applied approximations, ready for further processing.


\fakepar{Makefile generation}
\label{sec:llm-makefile}
\tool\ prompts the LLM to generate a makefile for automated compilation of the codebase. In this conversation (\textbf{Conv 3}, Figure~\ref{fig:archi}), a system prompt is issued \needo{[A.3.\ref{append:conv3sys}]}, followed by a user prompt listing the files in the codebase directory \needo{[A.3.\ref{append:conv3prompt1}]}. The LLM uses this information to produce the makefile, which incorporates all necessary compilation parameters, such as linker libraries and file names. The conversation history is saved for future reference, allowing for the generation of a new makefile if required in later steps.

\begin{lstlisting}[language=customc,caption={Approximate code with knob variables.}, label={approximated-snippet},  basicstyle=\small\ttfamily, numbers=none]
void sobel_filter(int* image) {
    /* Knob Variables Declaration Start */
    int knob1 = 80;
    /* Knob Variables Declaration End */
    image = load_image(IMAGE_SIZE);
    int* new_image = [IMAGE_SIZE];
    
    for (int i = 0; i < (IMAGE_SIZE*knob1)/100; i++) {
        /* Applying Sobel Filter */
    }
    send_to_classifier(new_image);
}
\end{lstlisting}

\begin{lstlisting}[language=customjson, caption={LLM's output in JSON }, label={lst:json_example}, basicstyle=\small\ttfamily, numbers=none, breaklines=true, breakatwhitespace=false, columns=flexible, frame=single, linewidth=\columnwidth]
{
    apx_code: "/*approximated code*/",
    knob_variables: ["knob1"],
    knob_ranges: [{"knob1": [20, 100]}],
    knob_increments: [{"knob1": "Integer"}]
}
\end{lstlisting}

\subsubsection{Code verification}

After a function is approximated, it is essential to verify that the modified code compiles and executes correctly. The approximated code is integrated into a copy of the codebase and prepared for testing.

\fakepar{Compilation}  
\tool\ compiles the patched codebase using the generated makefile. If compilation errors occur, the errors are logged and fed back to the LLM, which adjusts the makefile accordingly. This iterative process is repeated for a predefined number of attempts. If the code fails to compile after multiple iterations, \tool\ reverts to the beginning and applies an alternative approximation technique. Once the function successfully compiles, \tool\ transitions to the validation phase to evaluate runtime correctness.
The automated generation of makefiles and iterative compilation significantly reduces the developer’s workload, ensuring seamless integration of approximated code into the application’s build process. 

\fakepar{Validating Approximations}  
The final step in ensuring a safe approximation is runtime validation, which also confirms the safety of LLM-defined knob ranges. Validation starts by executing the approximated function using the upper and lower bounds of the knob values.
If the initial executions are error-free, \tool\ performs a binary traversal across the knob variable's range to identify specific intervals where runtime errors may occur. If errors are detected, the search is refined to focus on intervals where the function executes without issues. The approximation is deemed unsafe and discarded if the function fails across all tested intervals.

\begin{figure}
\centering
\includegraphics[width=\columnwidth]{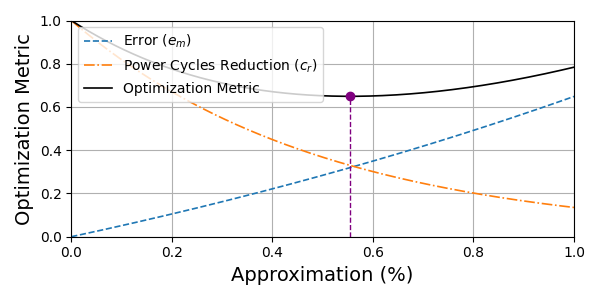}
\caption{Minimizing Optimization Metric: \textnormal{The purple dot indicates the point where the optimization metric achieves its minimum value, representing the balance between reduced power cycles and acceptable output error.}}
\label{fig:model}
\end{figure}

\subsection{Fine Tuning}
\label{sec:fine-tuning}

\fakepar{Optimization Function}  
After successfully validating all approximations, the next step is to determine the most suitable knob values that balance output error and the number of power cycles by minimizing the following optimization metric (see Figure~\ref{fig:model}):  
\begin{equation}
\label{eq:optimization-metric}
Optimization\ Metric = e_m + c_r
\end{equation}

where \( e_m \) quantifies the deviation between the original and approximated outputs:  
\begin{equation}
\label{eq:em}
e_m = \frac{\lvert a_o - a_a\rvert}{a_o}
\end{equation}  
where \( a_o \) and \( a_a \) denote the accuracy of the original and approximated outputs, respectively. \( e_m \) must remain below the error bound \( e_b \). The accuracy of the output is evaluated across four key output data-types currently supported by \tool, including numeric, text, image, and boolean (see Table~\ref{tab:accuracy-classes}), enabling broad applicability to IoT use cases like environmental sensing and edge machine learning.
Nonetheless, adding a new output type only requires adding a new error calculation function. 

The reduction in power cycles, \( c_r \), is defined as:  
\begin{equation}
\label{eq:checkpoint-reduction}
c_r = \frac{c_a}{c_o}
\end{equation}  
where \( c_o \) and \( c_a \) represent power cycles in the original and approximated programs, respectively. Typically, \( 0 < c_r \leq 1 \).

\tool\ employs Bayesian optimization~\cite{Sanders2019BayesianSF} to minimize the optimization metric leveraging a probabilistic model (e.g., Gaussian Process). Implemented using the \texttt{gp\_minimize} method from the scikit-optimize (\texttt{skopt}) library~\cite{tim_head_2018_1207017}, this approach ensures effective fine-tuning of knob variables. Traditional gradient-based methods, like gradient descent, are unsuitable due to the non-differentiable, computationally intensive nature of our objective function. Alternatives like brute-force and evolutionary algorithms are computationally prohibitive for time-sensitive evaluations.  

\fakepar{Simulation}  
Both $e_m$ and $c_r$ are evaluated using the \textit{Fused} simulator~\cite{Sliper2020Fused}, which models power consumption and energy storage, and emulates checkpointing and system reboots, vital for our evaluations. Fused was chosen over emulators like \textit{Renode}~\cite{renode} and \textit{MSPSim}~\cite{mspsim} for its extensive support for intermittent computing and compatibility with multiple architectures, including Cortex-M and MSP430.

\begin{figure*}[htbp]
    \centering
    \begin{minipage}{0.17\textwidth}
        \centering
        \includegraphics[width=\linewidth]{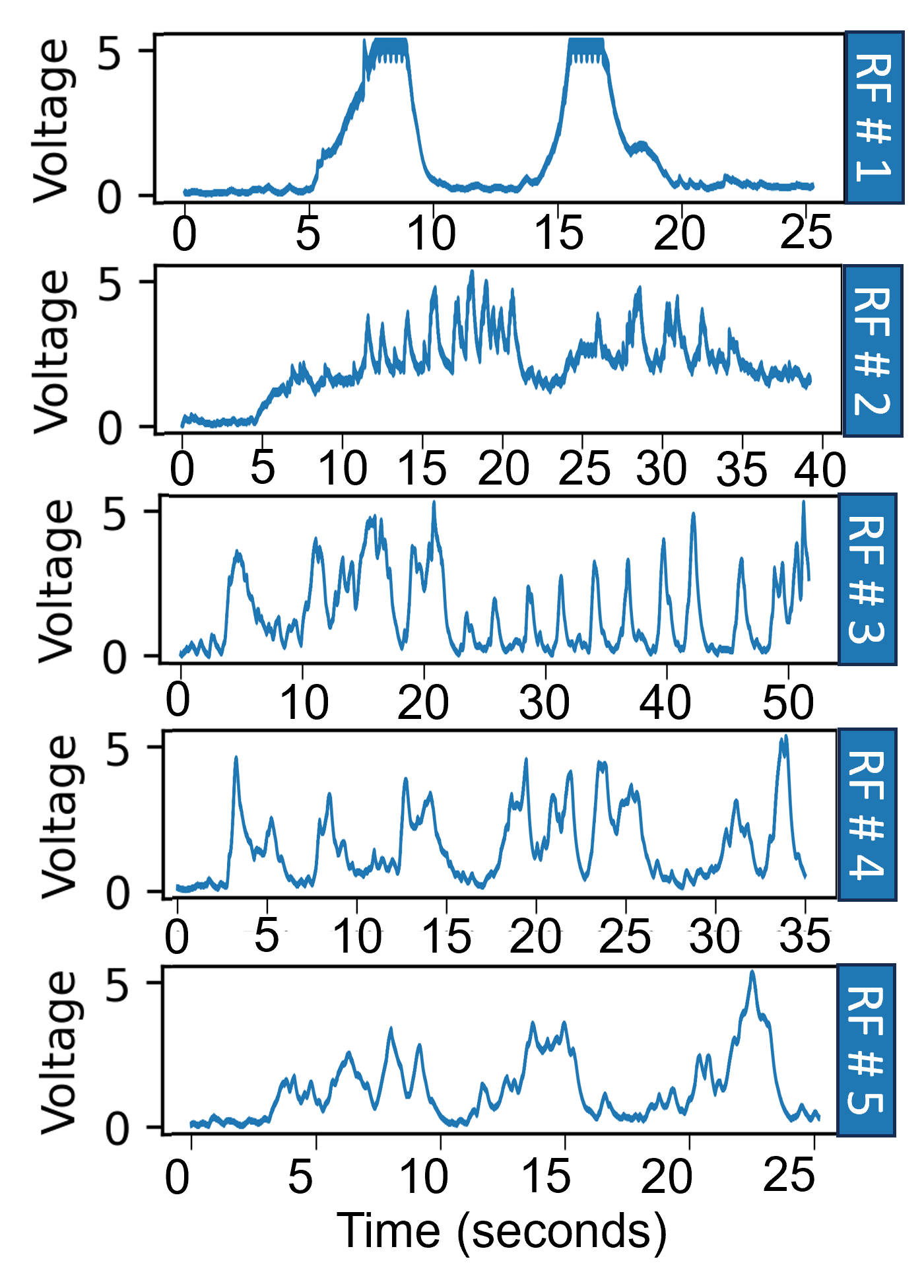}
        \caption{RF Traces}
        \label{fig:energy_traces}
    \end{minipage}
    \hfill 
    \hfill
    \begin{minipage}{0.8\textwidth}
        \centering
        \includegraphics[width=\linewidth]{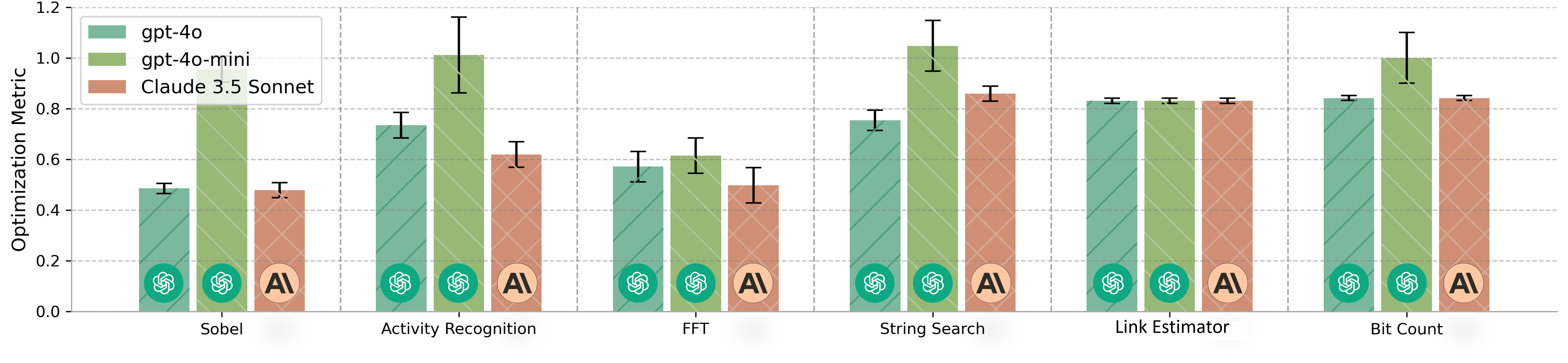}
        \caption{Optimization metrics for each LLM.\textnormal{Depicting performance of GPT-4o (right bar), GPT-4o Mini (middle bar), and Claude 3.5 Sonnet (left bar) across six applications. The lower values of the optimization metric ($e_m$+$c_r$) indicate better performance.}}
        \label{fig:llm_opt_metric}
    \end{minipage}
\end{figure*}

\section{Evaluation}
\label{sec:eval}


Our evaluation encompasses a detailed performance study, baseline comparisons, testbed experiments, and a small-scale user study.

\fakepar{Setting}  
For our evaluation, we employed ManagedState~\cite{sliper2019efficient} within Fused as the checkpointing mechanism for retaining program state across power cycles. Reducing checkpointing overhead, though, is an orthogonal problem, not the primary focus of our study. CheckMate's benefits are expected to generalize across various checkpointing solutions.

\begin{figure*}
    \centering
    \includegraphics[width=\textwidth]{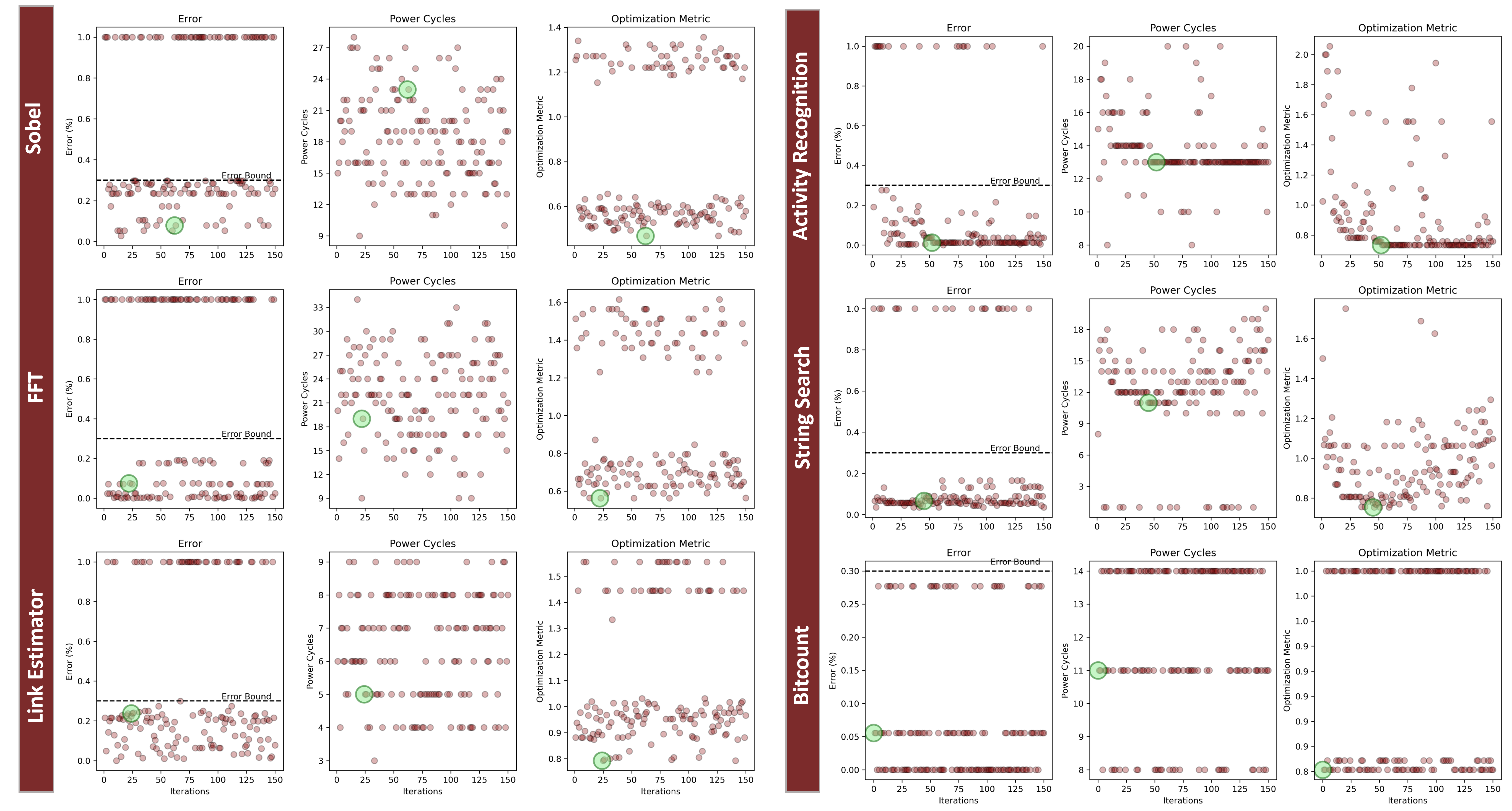}
    \caption{Bayesian optimization results for six applications. \textnormal{Each row represents an application, featuring subplots for output error (left), number of power cycles (middle), and optimization metric (right) across iterations. Green circles highlight the iteration at which the minimum optimization metric is reached, indicating the optimal balance between output error (\%) and the number of power cycles for each application.}}
    \label{fig:results}
\end{figure*}

For buffering harvested energy, our experiments used three capacitor sizes: the smallest required for each workload, as established in prior studies~\cite{colin2018reconfigurable,bhatti2017harvos}, and two subsequently larger standard sizes. To emulate realistic energy conditions, we incorporated five distinct open-source RF energy traces (Figure~\ref{fig:energy_traces}), commonly employed in previous research~\cite{energy-traces, bhatti2017harvos, mementos}.
We set a 30\% upper bound on output error, informed by prior research~\cite{dragos}, to define the search space for Bayesian optimization and enable effective fine-tuning of knobs. A user, however, can modify this value arbitrarily to align with their specific application requirements. The Bayesian optimization process,  conducted over 150 iterations as shown in Figure~\ref{fig:results}, balanced computational overhead with achieving solutions close to the global minimum. Further iterations provided only marginal gains in energy efficiency or accuracy, ensuring a practical trade-off between computation time and solution quality.

\fakepar{Applications}  
We evaluated \tool\ on six diverse applications, widely employed in intermittent computing evaluations~\cite{bhatti2017harvos,ahmed2019efficient,maeng2018adaptive,yildirim2018ink, mementos}, that offer diverse opportunities for approximation.

\begin{enumerate}[leftmargin=1.2em]
    \item \textit{\textbf{Sobel}}: An image processing algorithm for edge detection, highlighting regions of high spatial gradient. Commonly used in IoT applications such as real-time image processing on edge devices~\cite{joshi2022fast}. 
    \item \textit{\textbf{Activity Recognition}}: Classifies a user’s physical state (e.g., walking, running, stationary) using accelerometer data. Widely employed in IoT system benchmarks~\cite{ghibellini2022intelligence,rashid2022ahar}.
    \item \textit{\textbf{Fast-Fourier Transform (FFT)}}: Converts signals from the time to frequency domain, commonly used in IoT applications like real-time audio processing~\cite{antonini2018smart} and vibration analysis~\cite{verma2021edge}.
    \item \textit{\textbf{Boyer-Moore String Search}}: Efficiently locates substrings within a main string. Used in IoT applications for text processing and pattern recognition~\cite{domb2021framework, liu2022fast}.
    \item \textit{\textbf{Link Estimator}}: Evaluates the strength and reliability of wireless communication links, widely used by standard routing protocols like RPL for route establishment~\cite{rfc6550}.
    \item \textit{\textbf{Bitcount}}: Counts the number of set bits in a binary number, a key operation in IoT applications like data compression~\cite{sangodoyin2020side}.
\end{enumerate}

\begin{table}[tb]
             \centering
             \footnotesize
             \caption{Output types and accuracy classes.}
             \begin{tabular}{cc}
                 \toprule
                 \textbf{Program Output} & \textbf{Accuracy Class} \\
                 \midrule
                 \multirow{2}{*}{Numeric Values} 
                 & Raw Absolute Error\\
                 & Normalized R-squared\\
                 \midrule
                 \multirow{1}{*}{Text} & 1 - Word Error Rate\\
                 \midrule
                 \multirow{2}{*}{Image} 
                 & 1 - Pixel Error Rate \\
                 & Structural Similarity Index \\
                 \midrule
                 \multirow{1}{*}{Boolean} 
                 & F1-Score \\
                 \bottomrule
             \end{tabular}
             
             \label{tab:accuracy-classes}
 \end{table}

\begin{table}[tb]
    \centering
    \footnotesize
    \caption{\shepherdchange{\textbf{Application size}. \textnormal{The reported sizes exclude storage used by the checkpointing strategy, which can add up to 58 kB of Flash. Token counts represent the number of tokens required to tokenize the codebase.}}}  
    \begin{tabular}{lcccc}
        \toprule
        \textbf{Application} & \makecell{\textbf{Total Flash} \\ \textbf{Usage}} & \makecell{\textbf{Total RAM} \\ \textbf{Usage}} & \makecell{\textbf{Token} \\ \textbf{Count}}  & \makecell{\textbf{Line of} \\ \textbf{Code}} \\
        \midrule
        Sobel & 9.0 kB & 3712 B & 1002 & 123 \\ 
        \midrule
        Activity Recognition & 1200 B & 274 B & 2213 & 307 \\
        \midrule
        FFT & 11413 B & 26 B & 911 & 119 \\
        \midrule
        String Search & 4324 B & 942 B & 689 & 94 \\
        \midrule
        Link Estimator & 9600 B & 2516 B & 403 & 78 \\
        \midrule
        Bitcount & 574 B & 10 B & 489 & 80 \\
        \bottomrule
    \end{tabular}
    \label{tab:application-metrics}
\end{table}

\shepherdchange{As shown in Table \ref{tab:application-metrics}}, these benchmarks encompass a wide range of programming structures and varying code complexities, enabling a thorough evaluation of \tool's versatility and effectiveness across a spectrum of IoT applications. For instance, the Sobel filter involves computationally intensive image processing, while Activity Recognition employs classification with limited adjustable parameters. The FFT focuses on mathematical computations optimized for low-power microcontrollers, whereas the link estimator handles lightweight network statistics. The Boyer-Moore String Search addresses complex string matching with specific data dependencies, and Bitcount involves bit manipulation with simple control structures. 

\fakepar{Key findings} Our findings are:

\begin{itemize}[leftmargin=1em]
    \item Claude 3.5 Sonnet is identified as the most effective LLM for \tool, consistently delivering superior performance across applications while ensuring adaptability to future advancements in LLM technologies (Section~\ref{sec:llm_eval}).
    \item \tool demonstrates its effectiveness in optimizing energy efficiency across diverse IoT applications by achieving substantial reductions in power cycles (up to 60\%) while maintaining acceptable error margins in the output. (Section~\ref{sec:benchmarks}).
    \item \tool outperforms existing semi-automated frameworks like ACCEPT by achieving higher speedups and comparable or lower errors, all while operating fully autonomously, demonstrating its superiority and user-friendliness (Section~\ref{sec:comparative_analysis}).
    \item Evaluation on a testbed replaying energy harvesting traces and running the program on hardware demonstrates consistent performance, validating our simulation results (Section~\ref{sec:testbed}).

    \item A user study comprising 17 participants with varying programming expertise levels successfully used CheckMate to approximate IoT applications, highlighting its accessibility, ease of use, and potential for adoption in diverse user scenarios without requiring specialized knowledge (Section~\ref{sec:user_study}).

\end{itemize}


\subsection{Results \textrightarrow\ LLM Evaluation}
\label{sec:llm_eval}

As a preliminary step, we identified the most suitable LLM for \tool, as it directly affects the quality of approximations. We evaluated three commercially available LLMs: OpenAI’s GPT-4o (snapshot 2024-08-06), GPT-4o Mini (snapshot 2024-07-18), and Anthropic's Claude 3.5 Sonnet (snapshot 2024-10-22). Each model was rigorously tested across six applications, with experiments repeated 20 times to ensure statistical robustness. The aggregated results, including error bars representing confidence intervals, are shown in Figure~\ref{fig:llm_opt_metric}.

Our analysis shows that Claude 3.5 Sonnet consistently outperforms the other models, particularly in larger applications such as FFT. GPT-4o Mini exhibits greater variability, likely due to its limited computational resources, which can lead to suboptimal approximations. Based on these findings, we selected Claude 3.5 Sonnet as the primary LLM for \tool, ensuring reliable and accurate approximations in subsequent evaluations. As advancements in LLMs continue at a rapid pace, \tool is compatible with emerging models.

\subsection{Results \textrightarrow\ Performance}
\label{sec:benchmarks}
We conducted comprehensive performance benchmarks to evaluate \tool’s effectiveness under varying energy constraints. 


\begin{figure}
    \centering
    \includegraphics[width=\columnwidth]{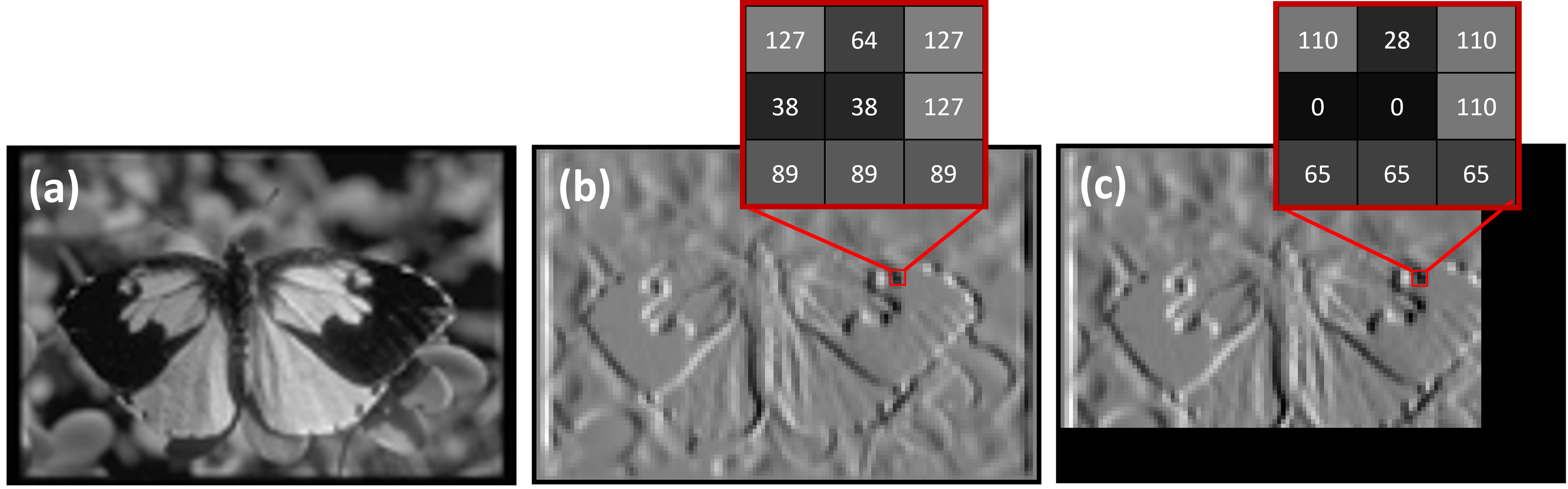}
        \caption{Original vs approximated Sobel. \textnormal{(a) The original image. (b) Edge detection using the original code. (c) Edge detection using the approximated code.}}
    \label{fig:susan_image_comparison}
\end{figure}

\begin{figure*}
    \centering
    \begin{subfigure}[b]{0.16\textwidth}
        \centering
        \includegraphics[width=\columnwidth]{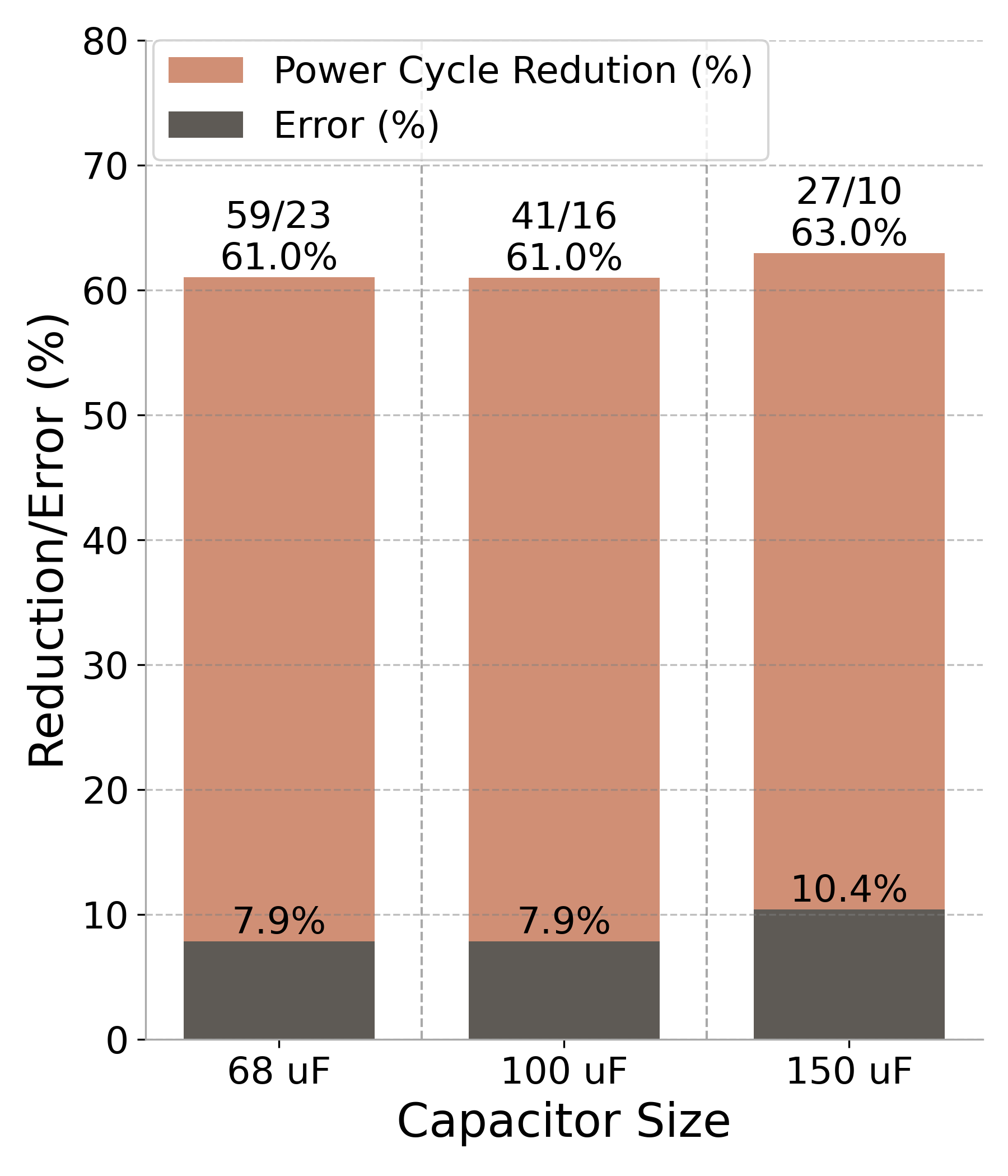}
        \caption{Sobel}
        \label{fig:sobel-cap-eval}
    \end{subfigure}
    \hfill
    \begin{subfigure}[b]{0.16\textwidth}
        \centering
        \includegraphics[width=\columnwidth]{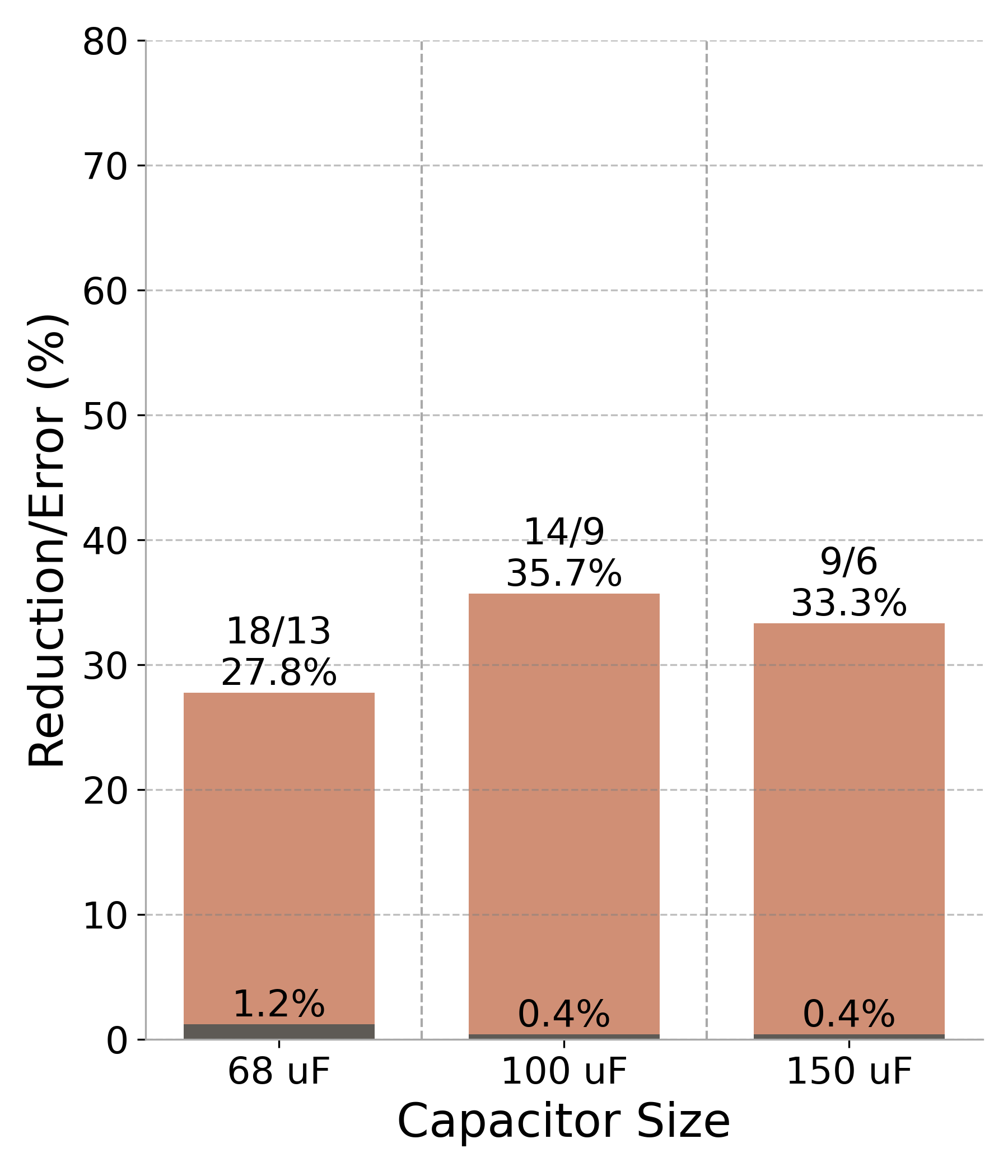}
        \caption{Activity Rec.}
        \label{fig:ar-cap-eval}
    \end{subfigure}
    \hfill
    \begin{subfigure}[b]{0.16\textwidth}
        \centering
        \includegraphics[width=\columnwidth]{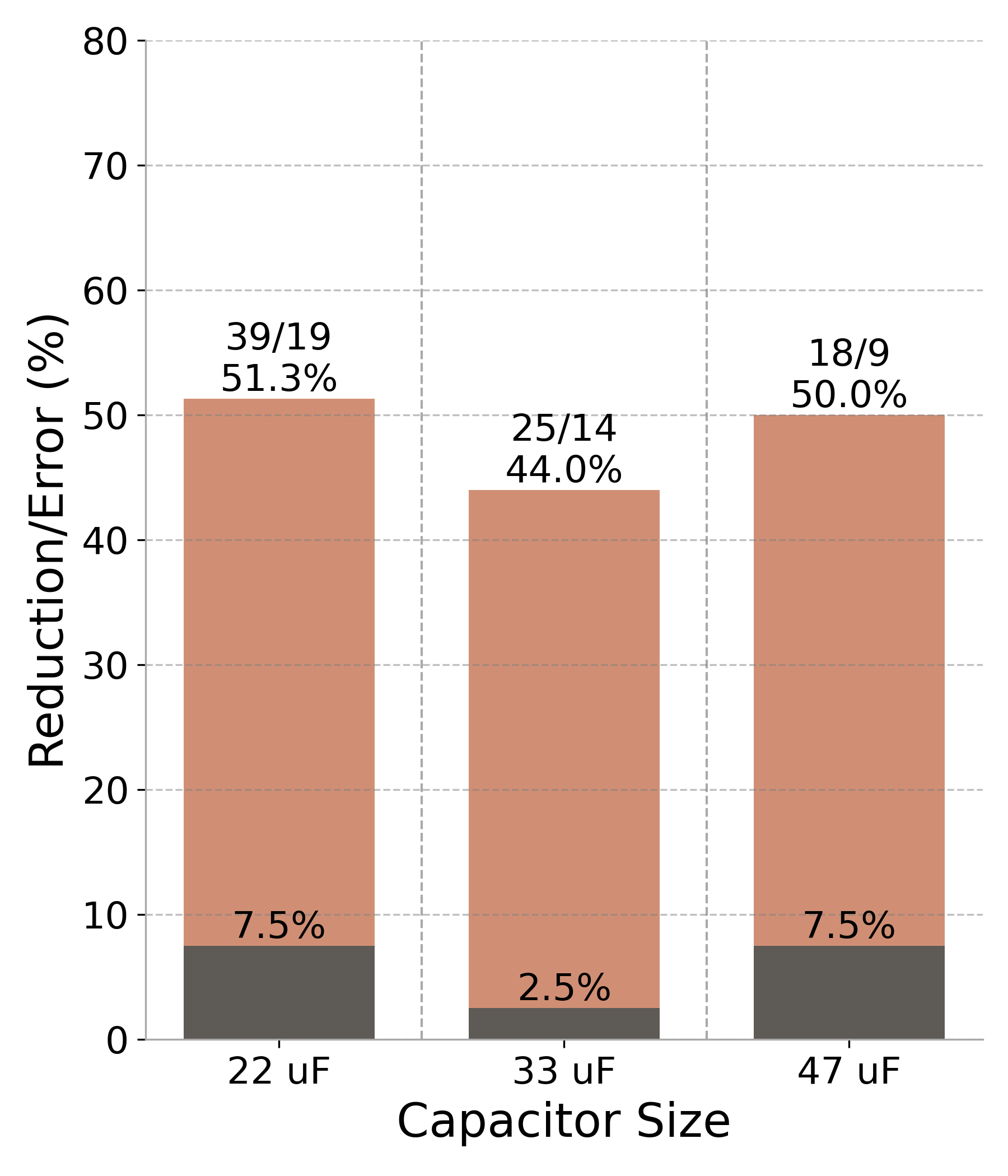}
        \caption{FFT}
        \label{fig:fft-cap-eval}
    \end{subfigure}
    \hfill
    \begin{subfigure}[b]{0.16\textwidth}
        \centering
        \includegraphics[width=\columnwidth]{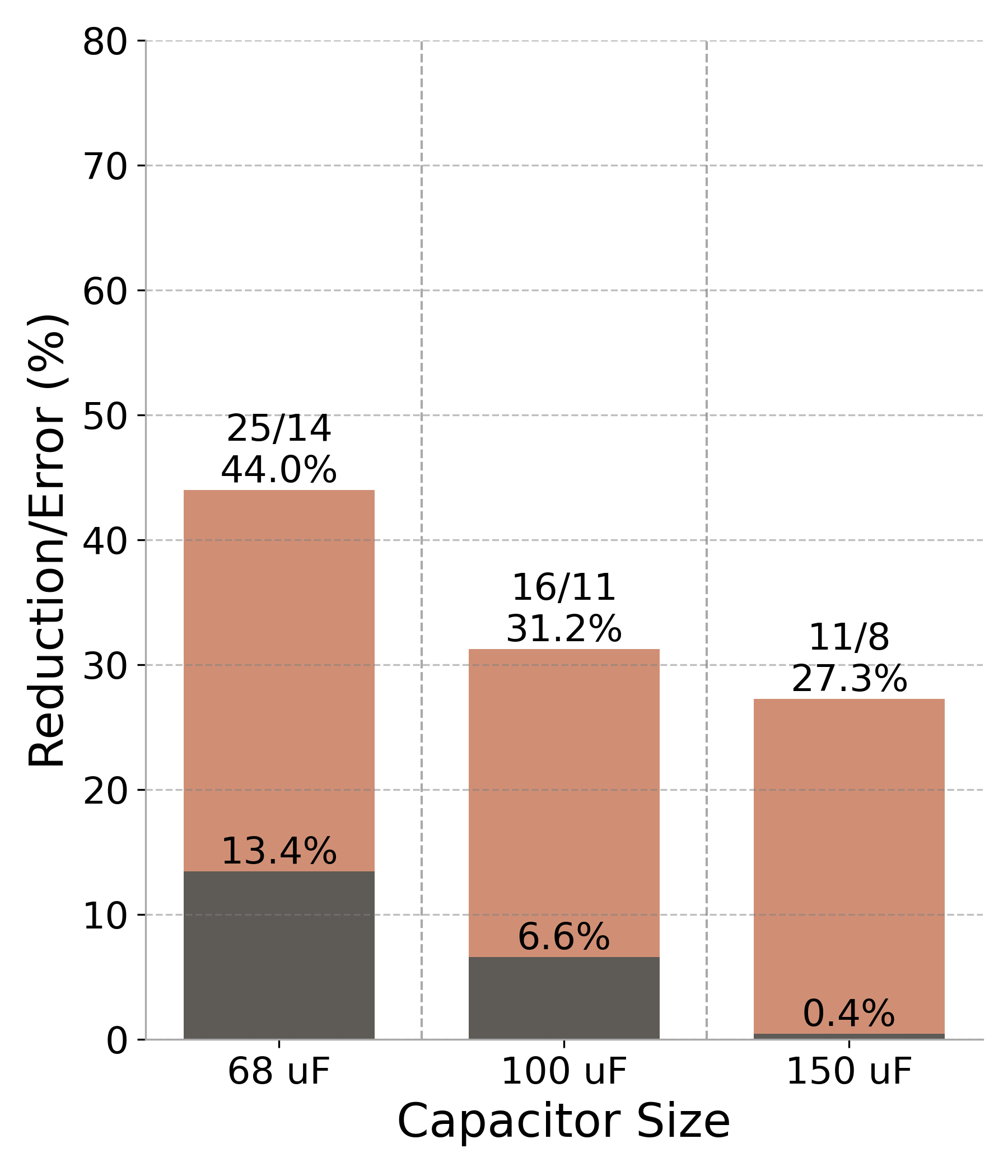}
        \caption{String Search}
        \label{fig:ss-cap-eval}
    \end{subfigure}
    \hfill
    \begin{subfigure}[b]{0.16\textwidth}
        \centering
        \includegraphics[width=\columnwidth]{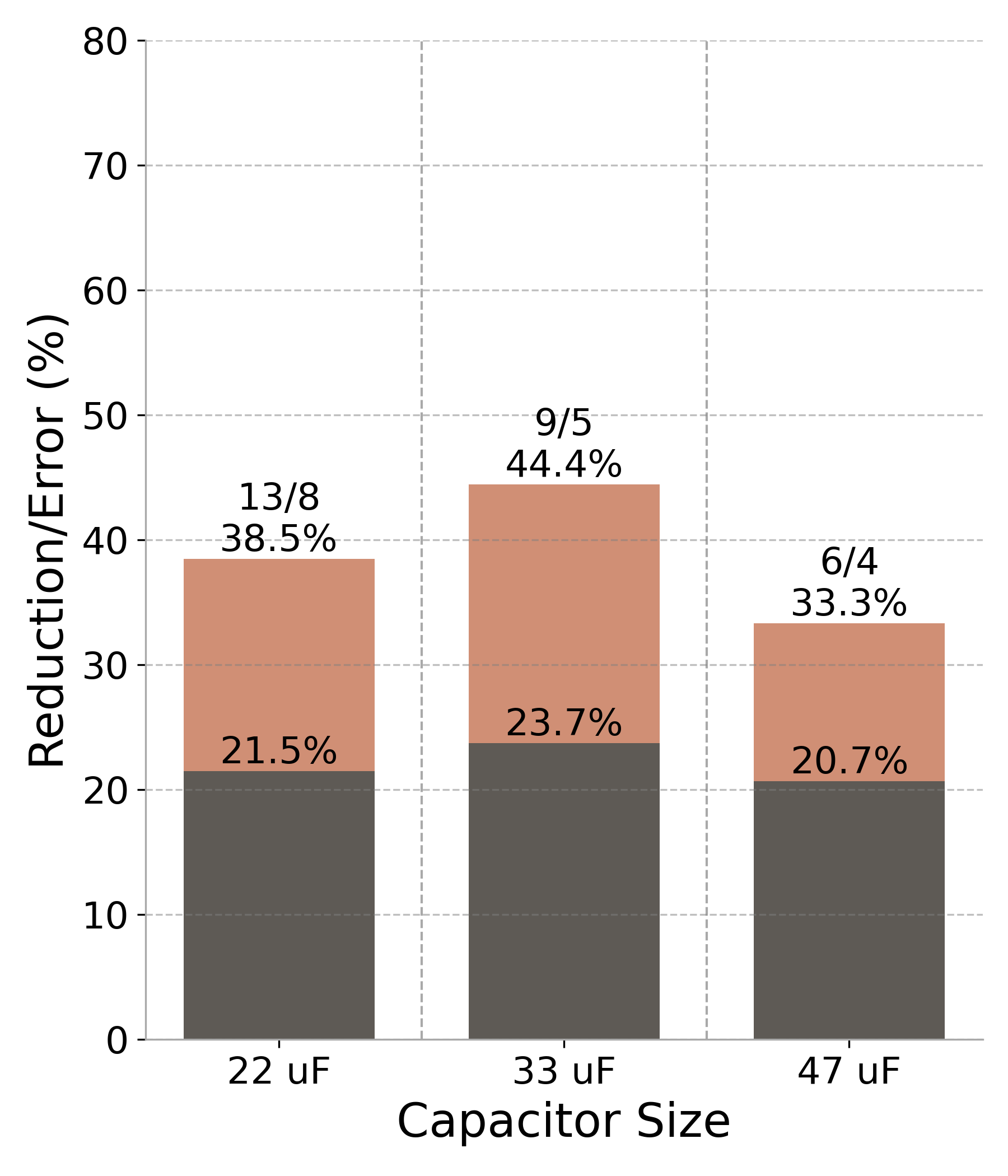}
        \caption{Link Estimator}
        \label{fig:lqi-cap-eval}
    \end{subfigure}
    \hfill
    \begin{subfigure}[b]{0.16\textwidth}
        \centering
        \includegraphics[width=\columnwidth]{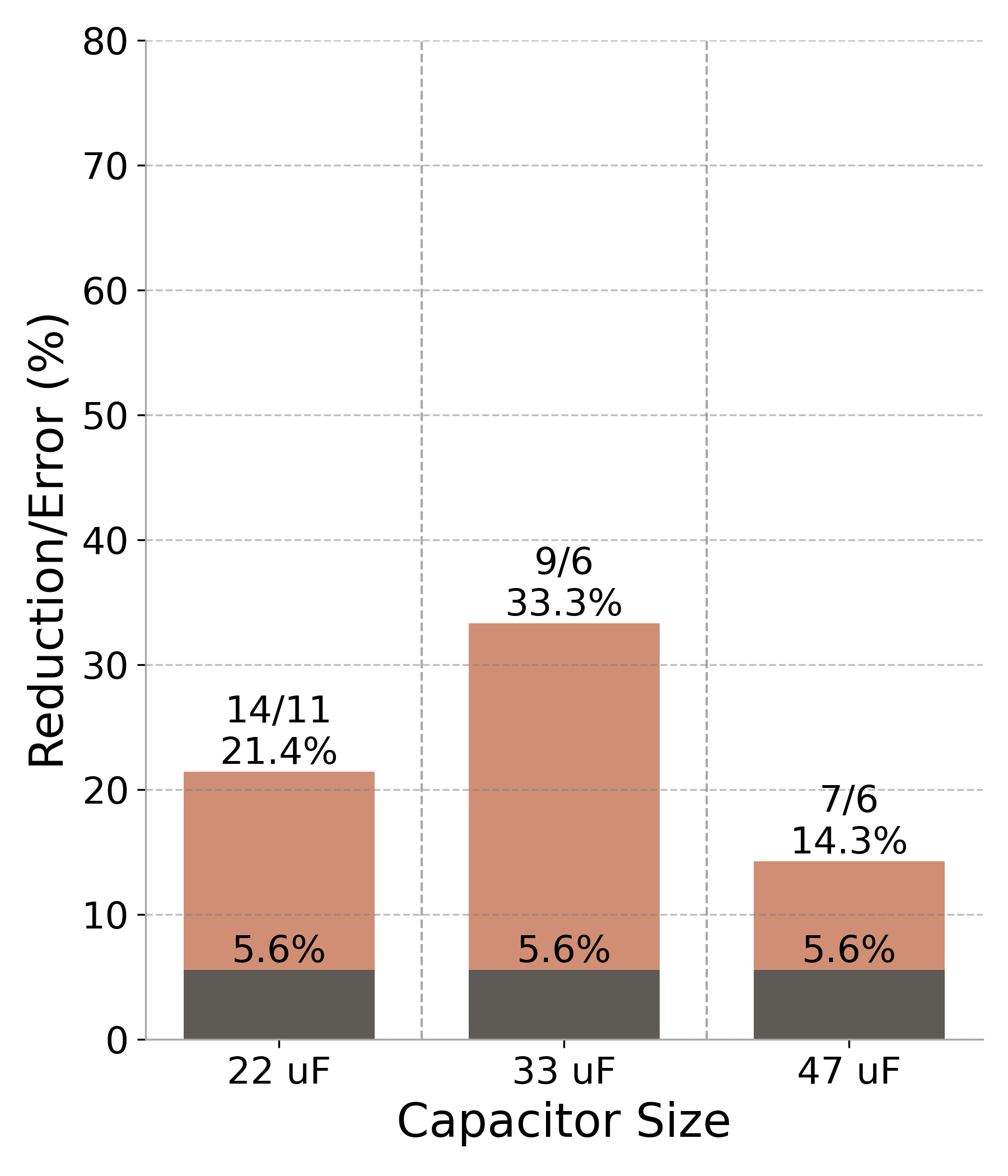}
        \caption{Bitcount}
        \label{fig:bitcount-cap-eval}
    \end{subfigure}

    \caption{Performance results (RF\#2 trace across three capacitor sizes). \textnormal{The darker bars represent the output error percentage, while the lighter bars illustrate the reduction in power cycles. Values above each bar indicate the original and approximated cycle counts, along with the corresponding percentage reduction.}}
    \label{fig:checkmate_evaluation}
\end{figure*}

\fakepar{Sobel filter} We evaluated the Sobel filter implementation from the ACCEPT repository~\cite{accept}, using the Structural Similarity Index Measure (SSIM), categorized under the image output type in Table~\ref{tab:accuracy-classes}, to assess output fidelity. \tool\ automatically identified two approximation opportunities in the Sobel filter: precision scaling and loop perforation. Precision scaling reduced numerical precision from double-precision floating-point to integer values, while loop perforation was applied to the input image and filtering loops.

During optimization, the Sobel filter’s knobs demonstrated high sensitivity, with only a narrow range yielding acceptable output errors. This sensitivity resulted in multiple iterations yielding a 100\% error, as any error surpassing the predefined threshold was capped at the maximum allowable value. Figure~\ref{fig:results} illustrates this, showing a distinct gap between two clusters in the optimization metric: the upper cluster reflects unacceptable errors, while the lower cluster includes iterations meeting the error criteria.

Despite these challenges, the optimization process delivered substantial improvements, , as shown in Figure~\ref{fig:sobel-cap-eval}. Precision scaling alone reduced power cycles by 49\% with minimal SSIM loss. Combining precision scaling with loop perforation further decreased power cycles from 59 to 23, achieving an overall reduction of approximately 60\%. These results were obtained with the smallest capacitor size of $68~\mu\text{F}$, the minimum required for Sobel filter operation. Similar power cycle reductions and output error rates were observed with larger capacitor sizes.
The visual effects of these approximations are depicted in Figure~\ref{fig:susan_image_comparison}. Precision scaling retains less information, causing pixels to shift to darker values due to approximate calculations. With input image truncation from loop perforation, two black bars appear, marking sections where the filter was not applied.

\fakepar{Activity Recognition} We use ACCEPT's implementation of activity recognition~\cite{accept}, using the F1-score as the primary metric to assess classification accuracy, chosen for its balanced consideration of precision and recall.

Through automated analysis, \tool\ identified two effective approximation techniques: loop perforation and feature approximation. Loop perforation involved selectively skipping certain data points to reduce computational load, while feature approximation simplified calculations of key statistical features, such as mean and standard deviation, during feature extraction. Combining these techniques achieved a 28\% reduction in power cycles (from 14 to 9) while maintaining high classification accuracy (see Figure~\ref{fig:ar-cap-eval}). 

The optimization process converged quickly, reaching its lowest metric at the 53rd iteration with an output error of only 1\%, attributed to the limited number of adjustable parameters. Figure~\ref{fig:results} illustrates the fine-tuning of activity recognition during, demonstrating the effectiveness of our approximation strategies.

\fakepar{FFT} We evaluated the Cooley-Tukey FFT algorithm~\cite{fft-implementation} using Root Mean Square Error (RMSE) as the accuracy metric, and minimum capacitor size of $22~\mu\text{F}$.

\tool\ identified five approximation strategies: Beyond optimizing the FFT algorithm itself, the LLM identified approximation opportunities in the Taylor series expansions for exponential and trigonometric functions, frequently used in FFT calculations. By adjusting the iteration count (i.e., the $n$-th term) in the Taylor series for $\cos$, $\sin$, and $\exp$ functions—and assigning a knob to control each—\tool\ managed to achieve a trade-off between output error and power cycle reduction. Precision scaling was also applied, switching from double-precision to single-precision floating-point data types.
These techniques achieved a 51\% reduction in power cycles with merely 7.5\% RMSE, as shown in Figure~\ref{fig:fft-cap-eval}.

\begin{table*}[ht]
\centering
\tiny
\caption{Reduction in power cycles(\%) and output error rates (\%) on remaining RF energy traces.}
\label{tab:multi-RF}
\resizebox{\textwidth}{!}{%
\begin{tabular}{
    l
    S[table-format=2.2]
    S[table-format=2.2]
    S[table-format=2.2]
    S[table-format=2.2]
    S[table-format=2.2]
    S[table-format=2.2]
    S[table-format=2.2]
    S[table-format=2.2]
    S[table-format=2.2]
    S[table-format=2.2]
    S[table-format=2.2]
    S[table-format=2.2]
}
\toprule
\multicolumn{1}{c}{\textbf{Trace}} &
\multicolumn{2}{c}{\textbf{Sobel}} &
\multicolumn{2}{c}{\textbf{AR}} &
\multicolumn{2}{c}{\textbf{FFT}} &
\multicolumn{2}{c}{\textbf{String Search}} &
\multicolumn{2}{c}{\textbf{Link Estimator}} &
\multicolumn{2}{c}{\textbf{Bit Count}} \\
\cmidrule(lr){2-3} \cmidrule(lr){4-5} \cmidrule(lr){6-7} \cmidrule(lr){8-9} \cmidrule(lr){10-11} \cmidrule(lr){12-13}
 & {Red. (\%)} & {Err. (\%)} &
 {Red. (\%)} & {Err. (\%)} &
 {Red. (\%)} & {Err. (\%)} &
 {Red. (\%)} & {Err. (\%)} &
 {Red. (\%)} & {Err. (\%)} &
 {Red. (\%)} & {Err. (\%)} \\
\midrule
RF \#1 & 63.16 & 17.37 & 40.10 & 4.87 & 47.67 & 0.01 & 33.33 & 2.48 & 36.36 & 21.50 & 36.36 & 27.72 \\
RF \#3 & 55.77 & 7.25  & 28.57 & 3.25 & 58.00 & 1.03 & 24.00 & 0.00 & 45.45 & 24.50 & 18.18 & 5.56  \\
RF \#4 & 54.00 & 11.92 & 35.29 & 3.25 & 51.23 & 0.06 & 25.00 & 2.48 & 41.67 & 21.50 & 41.67 & 27.72 \\
RF \#5 & 56.52 & 17.37 & 23.08 & 8.12 & 52.33 & 0.01 & 25.00 & 0.00 & 38.46 & 21.50 & 21.43 & 5.56  \\
\bottomrule
\end{tabular}%
}
\end{table*}

\fakepar{Boyer-Moore String Search} We employed Boyer-Moore string search algorithm from the MiBench2 benchmark suite~\cite{MiBench} using F1-score as the accuracy metric and a capacitor size of $68~\mu\text{F}$.

\tool\ achieved a 31\% reduction in power cycles with only a 6.6\% error in the output value on the smallest capacitor, as shown in Figure~\ref{fig:ss-cap-eval}. Importantly, this application demonstrates how LLMs can effectively interpret application context to apply domain-specific optimizations. During the function selection phase, the LLM identified an opportunity to approximate the initialization step in the Boyer-Moore algorithm. Specifically, it observed that the input traces did not cover the full ASCII range, reducing the \texttt{bad character table} to include only frequently encountered characters. 

We performed an additional test to validate this observation by withholding the input trace from the LLM. Without the input trace, the LLM skipped the \texttt{bad character table} approximation and relied solely on perforating the main search loop, resulting in a consistent 6.6\% error and a 10\% reduction in power cycles. This experiment clarified the consistently low error percentages seen during the fine-tuning process  (Figure~\ref{fig:results}), as adjusting the knob that controlled the \texttt{bad character table}'s size did not increase error but impacted power cycles due to the larger table’s storage and checkpointing requirements.
These findings underscore \tool's capacity to harness LLMs for intelligent, context-sensitive optimizations.

\fakepar{Link Estimator} We also evaluated \tool\ on a simple link quality estimator that applies an exponentially weighted moving (EWMA) average to packet reception rates. This estimator is an integral part of several RPL implementations~\cite{oikonomou2022contiki}

We evaluated the accuracy of the estimator using percentage absolute error. Given its straightforward codebase, this application offered a single approximation opportunity: loop perforation on link’s running history of packet reception rates. 
As shown in Figure~\ref{fig:lqi-cap-eval}, \tool\ achieved a 38\% reduction in power cycles with a 21\% error EWMA. Like the Sobel filter, the fine-tuning in Figure~\ref{fig:results} displayed two clusters: one with errors beyond the acceptable threshold and the other within acceptable bounds. Unlike Sobel and other applications, however, this application exhibited a narrower range of power cycles, likely due to its simple computations.

\fakepar{Bitcount} We adapted the bitcount implementation from the MiBench2 suite~\cite{MiBench} to evaluate \tool using 100 predefined random numbers. 
Due to the simplicity of this application, \tool\ identified only one viable approximation: perforating the main counting loop to include only the $n$ most significant bits. This approximation also had a small search space, resulting in a low-variance distribution observed in Figure~\ref{fig:results}. \tool\ achieved a 21.4\% reduction in power cycles with an output error of less than 6\% (Figure~\ref{fig:bitcount-cap-eval}).

\fakepar{Results summary}
Table~\ref{tab:multi-RF} summarizes our results across the remaining RF energy traces for the smallest capacitor size demonstrating that \tool retains its performance benefits across all traces and applications. Minor variations arise due to differing energy availability patterns in each trace, influencing the number of power cycles and subsequently affecting the optimization metric.

Benchmarking \tool\ across six diverse applications demonstrates its capability to generate energy-efficient, approximated code with minimal loss in output quality. Among the evaluated applications, the Sobel filter exhibited the most significant performance gains, underscoring \tool's effectiveness in optimizing resource-intensive image processing tasks. The activity recognition algorithm showcased rapid convergence with minimal error, highlighting efficiency in systems with limited adjustable parameters. The FFT evaluation illustrates \tool's strength in optimizing computationally intensive mathematical functions, achieving substantial energy savings. The link estimation algorithm exemplifies \tool's adaptability to lightweight applications with small power budgets. Lastly, the string search task emphasizes the value of context-aware approximation, as \tool\ leverages input characteristics to enhance efficiency. 

\begin{table}[t]
    \centering
    \scriptsize
    \caption{CheckMate vs ACCEPT}
    \resizebox{\columnwidth}{!}{
    \begin{tabular}{lllll}
    \toprule
    \multirow{2}{*}{Applications} & \multicolumn{2}{l}{ACCEPT} & \multicolumn{2}{l}{CheckMate} \\ \cmidrule(l){2-5} 
                                  & Speedup       & Error      & Speedup        & Error        \\ \midrule
    Sobel                         & 2.0$\times$          & $\approx$ 26.7\%       & \cellcolor{green!40} 2.6$\times$           & \cellcolor{green} $\approx$ 8.7\%          \\
    Activity Recognition          & 1.5$\times$          & $\approx$ 0.1 \%       & \cellcolor{green!20} 1.6$\times$            & \cellcolor{red!20} $\approx$ 0.6\%          \\ \bottomrule
    \end{tabular}}
    \label{tab:accept-compare} 
\end{table}

\subsection{Results \textrightarrow\ Baseline Comparison}
\label{sec:comparative_analysis}

To evaluate \tool's performance relative to existing frameworks, we compared it with ACCEPT~\cite{accept}, a semi-automated tool requiring expert knowledge for identifying approximation opportunities within the code. In contrast, \tool\ is fully automated, requiring no specialized user input.

We tested \tool\ on the two embedded applications, Sobel and Activity Recognition, from the ACCEPT repository, using the speedup metric—defined as the ratio of clock cycles in the original to approximated code—to measure processing time reduction, ensuring comparability with ACCEPT's results.

\shepherdchange{We selected these two applications to ensure a direct and fair comparison with ACCEPT. The developers of ACCEPT evaluated and reported results exclusively on these two embedded system applications, making them the only available baseline for comparison. Expanding our benchmark set beyond these applications would have introduced methodological inconsistencies, reducing the validity of direct performance comparisons.}

As shown in Table~\ref{tab:accept-compare}, \tool\ achieved significantly higher speedup for the Sobel application while maintaining greater accuracy. For Activity Recognition, \tool\ delivered comparable performance with a slightly higher error rate without any user intervention. These results demonstrate that \tool\ can match or surpass the efficiency of semi-automated, expert-driven frameworks while offering a single-click solution.

\subsection{Testbed Evaluation}
\label{sec:testbed}
\begin{figure}
    \centering
    \includegraphics[width=\columnwidth]{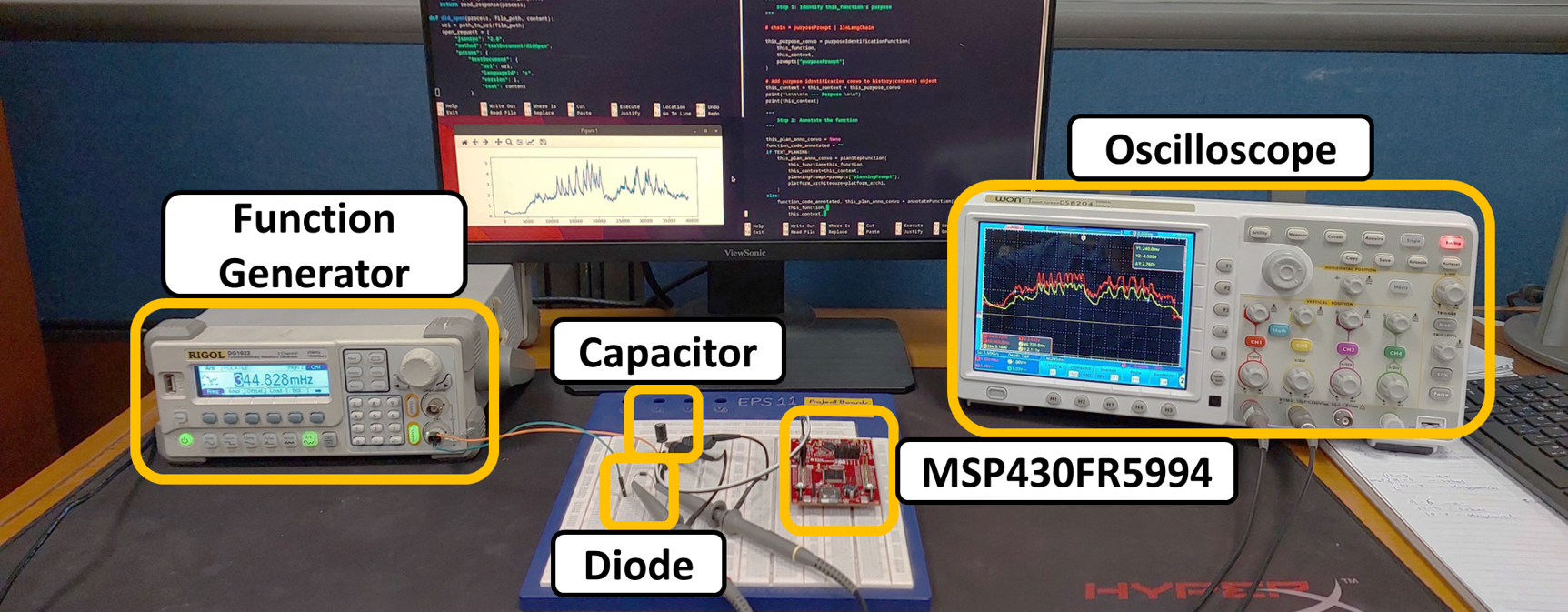}
    \caption{Testbed setup.}
    \label{fig:testbed}
\end{figure}

A testbed evaluation is crucial for assessing \tool's reliability beyond simulated environments.
Our testbed (Figure~\ref{fig:testbed}) consists of a function generator (RIGOL DG1022~\cite{RigolDG1022}) for replaying energy traces and the MSP430FR5994 Launchpad~\cite{ti_msp430} as the target platform.

The function generator, featuring a 14-bit digital-to-analog converter (DAC), scales voltage values in RF traces to integer values between 0 and $2^{14}$. This scaling is performed on a laptop and programmed into the generator using the Python \texttt{pyVISA} library~\cite{PyVISA}, which issues VISA commands for precise read-write operations. This approach eliminates the need for additional amplification circuitry, as the generator can directly supply sufficient current and voltage to the energy harvesting system.
The energy harvesting setup includes a capacitor, diode, and the MSP430. The capacitor stores energy from the generator's voltage signal, powering the MCU. The diode prevents reverse discharge, blocking current flow when the generator outputs a voltage lower than the capacitor’s stored charge. Voltage levels from both the energy trace and capacitor are monitored via an oscilloscope to ensure accurate energy trace replication.
To trace the power cycles required for program execution, we saved the count in a dedicated FRAM region on the MSP430.

We conducted experiments using five distinct RF energy traces~\cite{energy-traces} and present detailed results for the RF\#2 trace to compare these results with the ones presented in Figure~\ref{fig:checkmate_evaluation}. Table~\ref{tab:hardware-results} illustrates a strong correlation between simulation and real-world results.

We only report reductions in power cycles, as the error remains consistent for the approximated code in both simulation and hardware evaluations. Minor fluctuations in percentage reductions between the testbed and simulation results arise from real-world factors such as effective resistance, capacitor leakage, and other non-ideal parameters not fully modeled by the simulator. \needo{These discrepancies are more pronounced in applications with extended runtimes and larger codebases.}

\subsection{Small-scale user study}
\label{sec:user_study}

We conducted a user study involving seventeen computer science students to evaluate the user experience of \tool. Our key findings reveal that \tool{} offers two major benefits: 1) it significantly reduces the time required for performing approximations, and 2) it reduces the chances of higher errors in the output compared to manual methods.

\fakepar{Methodology} Participants were first introduced to approximate computing concepts and provided with documentation on writing approximate programs in C. They had 30 minutes to familiarize themselves with the material. To bridge theory with practice, each student was then walked through a complete application, showcasing effective application of approximation techniques. Following this, participants tackled three additional applications, identical to our benchmarks, applying approximation techniques over the course of one hour using standard C practices. Throughout these exercises, a cycle-accurate simulator was at their disposal, enabling real-time evaluation of their approximations. 

\begin{table}[t]
    \centering
    \scriptsize
    \caption{Simulation vs testbed}
    \resizebox{\columnwidth}{!}{%
    \begin{tabular}{l c c c}
    \toprule
    \textbf{Applications} & 
    \textbf{\shortstack{Sim Red.(\%)}} & 
    \textbf{\shortstack{Testbed Red.(\%)}} & 
    \textbf{\shortstack{Delta Diff.(\%)}} \\ 
    \midrule
    Sobel               & 61.0 & 66.6 & \textcolor{codegreen}{+5.6} \\
    Activity Rec & 35.7 & 33.3 & \textcolor{red}{-2.4} \\
    FFT & 51.3 & 49.4 & \textcolor{codegreen}{+1.9} \\
    String Search       & 31.2 & 30.0 & \textcolor{red}{-1.2} \\
    Link Estimator & 38.5 & 40.2 & \textcolor{red}{-1.7} \\
    Bit Count & 21.4 & 22.0 & \textcolor{red}{-0.6} \\
    \bottomrule
    \end{tabular}}
    \label{tab:hardware-results}
\end{table}

After completing the manual tasks, participants received documentation on \tool and a brief tutorial on its usage. They then reattempted the same programming challenges using \tool under identical conditions. Upon completion, participants filled out a survey assessing \tool's usability and effectiveness.

\fakepar{Participants} The study comprised computer science students from both junior and senior years of our undergrad program. These participants brought a diverse range of experiences, with three to seven years of formal computing education\footnote{Some students had extensively studied programming during high school.}. This varied background ensured a comprehensive evaluation of \tool across different academic levels and skill sets.

\fakepar{Results} 
Using \tool significantly reduced the average time required for performing approximations, decreasing from $13.93$ minutes (manual) to $2.56$ minutes per task. Moreover, manual approximations yielded suboptimal optimization metrics compared to \tool. Achieving a comparable optimization metric manually would require an indeterminate amount of time, making it impractical for evaluation within the study's timing constraints. Manual approximations produced output error rates ranging from 51\% to 82\% with inconsistent power cycle reductions\footnote{No correlation was observed between output error rates and power cycle reductions.}. In contrast, \tool consistently delivered superior results, as detailed in Section~\ref{sec:benchmarks}. Participants reported that \tool was intuitive and user-friendly, quickly mastering its features. A significant majority (93\%) indicated they would strongly recommend \tool for implementing approximations in intermittent computing applications.

\shepherdchange{
\section{Discussion and Limitations}
\label{sec:limitations}
The presented solution approach also introduces certain limitations and challenges related to the performance and practicability of \tool. In this section, we highlight some of these shortcomings and discuss potential solutions that users may adopt.

\fakepar{Approximation overhead} Several approximations introduce computational overhead. For example, a loop perforation approximation may add extra floating-point arithmetic operations, which contribute to the overall computation of the application. On its own, this overhead may not significantly alter the number of checkpoints required for the application to run to completion. However, when a large number of knobs are created by the LLM, it is possible for these knobs to collectively generate a significant overhead.

This overhead becomes particularly significant when many of the created knobs are `tuned out' by the Bayesian optimizer. 
A knob is considered tuned out when its setting ensures that the corresponding section of code behaves identically to the original application code. Consequently, tuned-out knobs introduce unnecessary computational overhead.

Much of this issue is mitigated by compilers, which automatically eliminate redundant code. However, there are instances where the compiler fails to remove such overhead. These typically arise when the LLM provides knob ranges \ref{lst:json_example} that lack a value capable of making the program behave exactly as the original code. In such cases, the Bayesian optimizer selects the knob value closest to "zero" (i.e., the setting most similar to the original program operation), but the compiler remains unable to eliminate the associated overhead.

\fakepar{Privacy concerns} Using commercial APIs for testing and evaluation of \tool can raise privacy concerns for developers operating in proprietary or confidential environments. To address this, \tool is designed to seamlessly integrate with any LLM API via LangChain, enabling it to operate with locally deployed LLMs. With the rapid advancements in open-source LLMs, such as DeepSeek's~\cite{deepseekv2}, many of which have reached performance levels comparable to commercial alternatives while requiring less computational power and budget, it is reasonable to assert that \tool would deliver similar performance when leveraging these open-source models.

\fakepar{LLM Specifications and Application Size Dynamics} The relationship between application size and the number of model parameters is a critical determinant of performance. As shown in Figure \ref{fig:llm_opt_metric}, \tool's effectiveness declines significantly when using smaller LLMs. For instance, while both GPT-4o and GPT-4o Mini share the same context window size, they differ in parameter count. The smaller GPT-4o Mini struggles with larger and more complex applications, exhibiting noticeable performance degradation. Additionally, it fails to capture some of the optimal approximations identified by its larger counterpart.  

Another key factor is the relationship between application size and context window length. Larger applications  require a longer context window for effective processing. To ensure broad generalizability across various LLMs while maintaining efficiency, \tool was designed to fit within a 16k-token context window\footnote{Different LLMs use varying tokenization strategies.}.

}

\section{Related Work}
Related efforts can be divided into two broad categories.

\fakepar{Application-specific approximations}
Approximation techniques in intermittent computing are often limited to specific applications. Bambusi et al.~\cite{bambusi2022case} introduced approximate intermittent computing, trading computational accuracy for reduced state maintenance overhead. Applied to human activity recognition and image processing, their approach achieved significant throughput gains with minimal accuracy loss.
In deep learning on intermittently powered devices, Islam and Nirjon~\cite{islam2019zygarde} developed Zygarde, which adjusts neural network computations based on energy availability for timely processing. Lin et al.~\cite{lin2023intermittent} proposed intermittent-aware neural network pruning, reducing computational load without significant loss in accuracy. Barjami et al.~\cite{lucaSenSys25} demonstrated that slight accuracy reductions can substantially boost throughput in intermittent inference, highlighting the potential of controlled approximation.  In \cite{energy_efficiency_paper}, the authors explore algorithmic-level approximation techniques such as loop perforation and synaptic pruning specifically for neural networks. 
Ganesan et al.~\cite{ganesan2019s} presented the "What's Next" architecture, integrating approximate computing into intermittent systems using anytime algorithms and hardware modifications. Evaluated on a 2D convolution application, it effectively maintained output quality under energy constraints.

\fakepar{Approximation frameworks}
 Closer to our work are solutions that provide frameworks for applying approximations. ACCEPT~\cite{accept} provides a comprehensive framework that integrates compiler analysis and auto-tuning to automate program relaxation while adhering to safety constraints. However, it requires extensive manual tuning and expert intervention to achieve acceptable results. Puppeteer~\cite{pup2022} allows developers to annotate code regions to quantify the sensitivity of application outputs to approximation errors. While this provides some control, it places a significant burden on developers to correctly identify and annotate sensitive code regions. Rumba~\cite{rumba} tackles the issue of managing error in approximations through continuous verifications and adjustments. Yet, it adopts a one-size-fits-all strategy that lacks the flexibility needed to balance energy efficiency and computational accuracy in highly diverse applications.

\tool\ is a software-only framework designed to address distinct challenges of intermittent computing on batteryless IoT. Unlike generic approximation tools, \tool\ uniquely balances power cycles and output errors, optimizing energy efficiency while ensuring desired output fidelity.

\section{Conclusion}
We presented \tool, a novel framework that leverages LLMs to automate approximate intermittent computing. \tool integrates LLM-driven approximations with robust validation processes, dynamic knobs, and Bayesian optimization to effectively balance energy efficiency and computational accuracy with minimal developer input. Evaluations across diverse IoT applications demonstrated substantial reductions in power cycles while maintaining output errors within user-defined bounds. Real-world testbed results aligned closely with simulation predictions, underscoring \tool's reliability. These advancements position \tool as a foundational step towards enabling scalable and user-friendly, automated approximation solutions for batteryless IoT environments.

\clearpage
\bibliographystyle{ACM-Reference-Format}
\bibliography{references}

\clearpage
\appendix
\shepherdchange{

\section{LLM Prompts Used in This Study}
\label{sec:appendix-llm-prompts}

This appendix provides the detailed LLM prompts used in our research study, divided into two main conversations: the "Approximation Engine" and the "Makefile Generator."

\subsection{Context Aware App Analyzer}

\begin{lstlisting} [label={append:conv1sys},caption=System prompt.]
You are "CheckMate," an LLM tool used for applying code approximations (approximate computing techniques). Our goal is to reduce program clock cycles to reduce energy consumption. For this, we are willing to accept some output errors. 
\end{lstlisting}

\begin{lstlisting} [label={append:conv1prompt1},caption=Generate a summary of the code base.]
Here is the codebase of the application:

{complete_code_base}
    
Create a small summary of the app and its purpose. Then, for each function, write a brief summary of its purpose.
\end{lstlisting}

\begin{lstlisting} [label={append:conv1prompt2},caption=Detailed function discription.]

Let's discuss {function_name} function.

What is the purpose of {function_name} function? 
What are its inputs and outputs? How does it accomplish its purpose? 
How would its output interact with the rest of the code? 

\end{lstlisting}

\begin{lstlisting} [label={append:conv1prompt3},caption=Function selection prompt.]
Now, identify which functions should be approximated. For all functions in the code, please reply in this format:

# Function Name

Resoning for approximating or not approximating function.

# List of functions
{
  "function_1": "approximate",
  "function_2": "do not approximate",
  ...
  (all functions in code)
}

\end{lstlisting}

\subsection{Approximation Engine}

\begin{lstlisting} [label={append:conv2sys},caption=System prompt.]
You are "CheckMate", an LLM tool designed to apply code approximations-techniques that intentionally introduce minor computational errors to reduce clock cycles, leveraging tolerance for inaccuracy in certain applications. The applications you are dealing with relate to batteryless IoT devices, where reducing power consumption and optimizing for low-energy environments is critical.

We will follow this flow to apply approximations:

    1. Planning/Annotation Step: After determining the function's purpose, you will assess whether it is safe to approximate. If so, you will annotate the code, adding comments where approximations can be applied, along with descriptions of what and how to apply them.

    2. Approximation Step: You will then be prompted again to apply the approximations you previously annotated.

These steps will be repeated for each function in our program's codebase. If a function contains calls to other functions, the conversation history of applied approximations for the called functions will be provided, enabling you to apply more effective approximations.

{code_base_summary}
\end{lstlisting}

\begin{lstlisting} [label={append:conv2prompt1},caption= Approximation identification.]
Now, let's move to the planning step. For the given function, {function_name}, what are the possible approximations that can be performed?

If there exists a knob variable (a variable that, when changed, increases or decreases the level of approximations), state that it is a candidate.

A point to note, this code is meant for {platform_architecture} architecture, so your suggested changes should be possible within the architecture.
\end{lstlisting}

\begin{lstlisting} [label={append:conv2prompt2},caption= Approximation implementation.]
Now, for the {function_name} function, approximate the code using the approximations highlighted in the previous prompt. The goal is to modify this function to reduce the number of clock cycles it uses. Focus on the identified areas of potential improvement and apply the described approximations. Additionally, for any of the knob variables identified or created, ensure they are declared at the top of the function within the /* Knob Variables Declaration Start */ and /* Knob Variables Declaration End */ comments.

Return the following:

    The approximated code.
    A list of knob variables.
    A list of ranges for each knob variable.
    The step size by which to increment each knob variable; two options: Real or Integer.
\end{lstlisting}

\begin{lstlisting} [label={append:conv2prompt3},caption= Approximation JSON conversion.]
{add_error}
Reformat the code you just generated into a JSON object. The JSON should contain the following information:

    1. approximated_code: The full block of code that you generated.
    2. knob_variables: A list of all the variable names that can be tuned or adjusted in the code. These are the "knobs."
    3. knob_ranges: The possible ranges of each knob variable (in the form of minimum and maximum values or specific values, if applicable).
    4. knob_increments: The increments by which each knob variable can change. Specify whether the increment is a Real number or an Integer.

Format the output as a clean and structured JSON object as described here:

{output_instuctions}
\end{lstlisting}

\begin{lstlisting} [label=append:fewshot-loop-perf, caption= Few-shot prompt to guide LLM efficient loop perforation.]
Here is an examples of loop perforation. If applying loop perforation only apply in this format. No other format of loop perforation is ever acceptable. Do not add in the iteration step, only the condition step.

#### Original un-perforated function

```c
void susan_edges(in, r, mid, bp, max_no, x_size, y_size)
    uchar *in,
    *bp, *mid;
int *r, max_no, x_size, y_size;
{
  /*Knob Variables Declaration Start*/
  int loop_skip = 2;
  float precision_scale = 0.9;
  /*Knob Variables Declaration End*/

  float z;
  int do_symmetry, i, j, m, n, a, b, x, y, w;
  uchar c, *p, *cp;

  memset(r, 0, x_size * y_size * sizeof(int));

  for (i = 3; i < y_size - 3; i += loop_skip) /* @Approximation applied [No.1] [Loop Perforation] */
    for (j = 3; j < x_size - 3; j += loop_skip) /* @Approximation applied [No.1] [Loop Perforation] */
    {
      n = 100;
      p = in + (i - 3) * x_size + j - 1;
      cp = bp + in[i * x_size + j];
    }
}  
```

#### Perforated function

```c
void susan_edges(in, r, mid, bp, max_no, x_size, y_size)
    uchar *in,
    *bp, *mid;
int *r, max_no, x_size, y_size;
{
  /*Knob Variables Declaration Start*/
  int loop_perforation_factor = 0.2;
  float precision_scale = 0.9;
  /*Knob Variables Declaration End*/

  float z;
  int do_symmetry, i, j, m, n, a, b, x, y, w;
  uchar c, *p, *cp;

  memset(r, 0, x_size * y_size * sizeof(int));

  int loop_truc1 = (y_size - 3) * loop_perforation_factor /* truncating the loop */
  int loop_truc2 = (y_size - 3) * loop_perforation_factor /* truncating the loop */
  for (i = 3; i < loop_truc1; i++) /* @Approximation applied [No.1] [Loop Perforation] */
    for (j = 3; j < loop_truc2; j++) /* @Approximation applied [No.1] [Loop Perforation] */
    {
      n = 100;
      p = in + (i - 3) * x_size + j - 1 - n;
      cp = bp + in[i * x_size + p];
    }

  return cp;
}
```
\end{lstlisting}

\begin{lstlisting} [label=append:fewshot-precision, caption= Precision scaling example to help LLM understant safe and effective way to precision scale.]
Here is an examples of precision scaling.
### Original unscaled function

```c
void calculate_image_gradient(int *gradient, float *image, int width, int height)
{
    int i, j;
    float gx, gy, magnitude;

    for (i = 1; i < height - 1; i++) /* Calculate gradients for all pixels */
        for (j = 1; j < width - 1; j++)
        {
            gx = (image[(i - 1) * width + (j + 1)] - image[(i - 1) * width + (j - 1)]) +
                 2 * (image[i * width + (j + 1)] - image[i * width + (j - 1)]) +
                 (image[(i + 1) * width + (j + 1)] - image[(i + 1) * width + (j - 1)]);
            gy = (image[(i - 1) * width + (j - 1)] + 2 * image[(i - 1) * width + j] + image[(i - 1) * width + (j + 1)]) -
                 (image[(i + 1) * width + (j - 1)] + 2 * image[(i + 1) * width + j] + image[(i + 1) * width + (j + 1)]);
            magnitude = sqrt(gx * gx + gy * gy);
            gradient[i * width + j] = (int)magnitude;
        }
}
```

---

### Scaled function with **Precision Scaling**

```c
void calculate_image_gradient(int *gradient, int *image, int width, int height)
{
    int i, j;
    int gx, gy, magnitude;

    for (i = 1; i < height - 1; i++) /* Calculate gradients for all pixels */
        for (j = 1; j < width - 1; j++)
        {
            gx = (image[(i - 1) * width + (j + 1)] - image[(i - 1) * width + (j - 1)]) +
                 2 * (image[i * width + (j + 1)] - image[i * width + (j - 1)]) +
                 (image[(i + 1) * width + (j + 1)] - image[(i + 1) * width + (j - 1)]);
            gy = (image[(i - 1) * width + (j - 1)] + 2 * image[(i - 1) * width + j] + image[(i - 1) * width + (j + 1)]) -
                 (image[(i + 1) * width + (j - 1)] + 2 * image[(i + 1) * width + j] + image[(i + 1) * width + (j + 1)]);
            magnitude = abs(gx) + abs(gy); /* Approximation by avoiding sqrt */
            gradient[i * width + j] = magnitude; /* Direct assignment as int */
        }
}
```

\end{lstlisting}

\subsection{Makefile Generator}

\begin{lstlisting}[label={append:conv3sys}, caption=System prompt.]
You are an AI assistant that generates Makefiles based on a given list of source files. A Makefile is a file used by the make utility to manage the build automation of projects. The goal is to compile the source code files into an executable or library.

Instructions:

    1. You will be provided with a list of source files.
    2. Generate a Makefile that compiles these files into an executable named main.
    3. Include the necessary rules to compile object files from the source files.
    4. Create a clean rule to remove all object files and the executable.
    5. Use gcc as the compiler.
\end{lstlisting}

\begin{lstlisting}[label={append:conv3prompt1}, caption=Makefile generation]
Given a list of files in a directory, output a Makefile to compile the application.

    files = {files_list}

    Output only the Makefile content, no other text. Your exact output will be pasted into the Makefile (so do not include "```"). The command that will be run is just "make main". Do not use the -Werror or -Wall flag."
\end{lstlisting}

}

\end{document}